\documentclass[aip,jap,preprint,groupedaddress]{revtex4-1}
\usepackage{graphicx}
\usepackage{hyperref}

\newcommand*\rfrac[2]{{}^{#1}\!/_{#2}}
\begin{document}
\title{Antiferromagnetism and superconductivity in the half-Heusler semimetal HoPdBi}
\author{Orest Pavlosiuk$^1$} \author{Dariusz Kaczorowski$^1$} \author{Xavier Fabreges$^2$} \author{Arsen Gukasov$^2$} \author{Piotr~Wi\'sniewski$^1$*}
\affiliation{$^1$\,Institute for Low Temperatures and Structure Research,
Polish Academy of Sciences, P. O. Box 1410, 50-950 Wroc{\l}aw Poland\\
$^2$\,L\'eon Brillouin Laboratory, CEA-CNRS, CE Saclay, 91191 Gif/Yvette France} 
\date{\today}
\begin{abstract}
We observed the coexistence of superconductivity and antiferromagnetic order in the single-crystalline ternary pnictide HoPdBi, a plausible topological semimetal. The compound orders antiferromagnetically at $T_{\rm N} = 1.9\,$K and exhibits superconductivity below $T_{\rm c} = 0.7\,$K, which was confirmed by magnetic, electrical transport and specific heat measurements. The specific heat shows anomalies corresponding to antiferromagnetic ordering transition and crystalline field effect, but not to superconducting transition. Single-crystal neutron diffraction indicates that the antiferromagnetic structure is characterized by the ($\rfrac{\,1}{2},\rfrac{1}{2},\rfrac{1}{2}$) propagation vector. Temperature variation of the electrical resistivity reveals two parallel conducting channels of semiconducting and metallic character. In weak magnetic fields, the magnetoresistance exhibits weak antilocalization effect, while in strong fields and temperatures below 50\,K it is large and negative. At temperatures below 7\,K Shubnikov-de Haas oscillations with two frequencies appear in the resistivity. These oscillations have non-trivial Berry phase, which is a distinguished feature of Dirac fermions.  \\\\\\\\\\\\\\\\~~* corresponding author
\end{abstract}
\maketitle
More than half a century ago it has been discovered that superconductivity (SC) and antiferromagnetism (AFM) can coexist in one compound, but new reports continue to appear describing this uncommon fusion of properties. This phenomenon can be observed most often in cuprates, iron-based pnictides and chalcogenides, and in a number of heavy fermion materials. Recently, SC and magnetic order were discovered in several half-Heusler compounds containing rare earths.\cite {Pan2013a, Goraus2013, Nakajima2015, Nikitin2015} This group consists of compounds crystallizing in the non-centrosymmetric MgAgAs-type structure. Excitingly, in some half-Heusler phases containing heavy elements (i.e. with strong spin-orbit coupling) an inversion of $\Gamma_8$ and $\Gamma_6$ bands occurs,\cite{Chadov2010a,Nowak2015} which is the necessary condition for non-trivial $Z_2$ topology.\cite{Ando2013, Hasan2010a}
Therefore half-Heusler phases are promising materials due to coexistence and interplay of AFM, SC and potential topological properties.

Rapidly expanding group of Dirac materials is characterized by low--energy electron behavior described by the relativistic Dirac equation.\cite{Vafek2014} New quantum states of matter such as topological semimetals, also called Dirac semimetals, have been identified later than other Dirac materials and were studied to lesser degree.
Similarly to topological insulators, 3D Dirac semimetals have inverted-band electronic structures but the difference between them is that the conduction and valence bands of Dirac semimetals contact only at so-called Dirac points, whereas topological insulators have bulk band gap in 3D momentum-space. Moreover topological semimetals host linear dispersions in all directions around Dirac nodes.\cite{Young2012} Such unusual properties of electron bands make materials of this class 3D-analogues of graphene and  surface of topological insulators, both with linear band dispersions in 2D momentum-plane.\cite{Castro2009, Ando2013, Hasan2010a} Besides, topological semimetals possess 'topologically protected' surface states containing Fermi arcs, that were theoretically predicted on the example of pyrochlore iridates\cite{Wan2011} and proved experimentally on Cd$_2$As$_3$ single crystals.\cite{Yi2014}

It has recently been proposed, based on {\it ab initio} electronic structure calculations, that half-Heusler HoPdBi can be a topological semimetal.\cite{Nikitin2015} The first investigations of $RE$PdBi systems pertained to magnetic and transport properties and were performed on polycrystalline samples.\cite{Riedemann1996, Gofryk2005, Gofryk2011, Kaczorowski2005} Almost all of them showed antiferromagnetic ordering at low temperatures and electrical transport measurements revealed a semimetallic or narrow-band-semiconducting nature.

The next stage of the research on half-Heusler phases has been associated with findings of the band inversion in some of them.\cite{Chadov2010a, Al-Sawai2010, Lin2010, Nowak2015} Observation of SC in some representatives of this family: LaPtBi ($T_{\rm c}$ = 0.9\,K),\cite{Goll2008a} LuPtBi (1.0\,K),\cite{Tafti2013} YPtBi (0.77\,K),\cite{Butch2011a, Bay2012} and LuPdBi (1.7-1.9\,K),\cite{Xu2014a, Pavlosiuk2015} further increased the interest in their physics, because the simultaneous observation of topological surface states and SC lays the groundwork for realization of Majorana fermion states on the surface  of topological superconductors.\cite{Fu2008}

The discovery of AFM in HoPdBi has opened a possibility of it being an antiferromagnetic topological insulator (AFTI).\cite{Gofryk2005} A theory of AFTI was proposed by Mong et.al,\cite{Mong2010} decribing how the time-reversal symmetry and the primitive-lattice translational symmetry are broken by magnetic ordering, whereas the symmetry being their product is retained. Magnetic structure with the ($\rfrac{\,1}{2},\rfrac{1}{2},\rfrac{1}{2}$) propagation vector is required for the occurence of topological states in a half-Heusler antiferromagnet. Such propagation vector has been determined for another antiferromagnetic half-Heusler compound GdPtBi.\cite{Muller2014} 

Here we report on magnetic, electrical transport, specific heat and neutron diffraction measurements performed on single crystals of HoPdBi. Comparison to data very recently reported in Refs.~\onlinecite{Nakajima2015} and \onlinecite{Nikitin2015} is made as well. Our results provide solid evidence for magnetic order at 1.9\,K and superconductivity at 0.7\,K. 
\section*{Results}
\subsection*{Electrical resistivity}
\begin{figure}[h]
\includegraphics[width=0.49\textwidth]{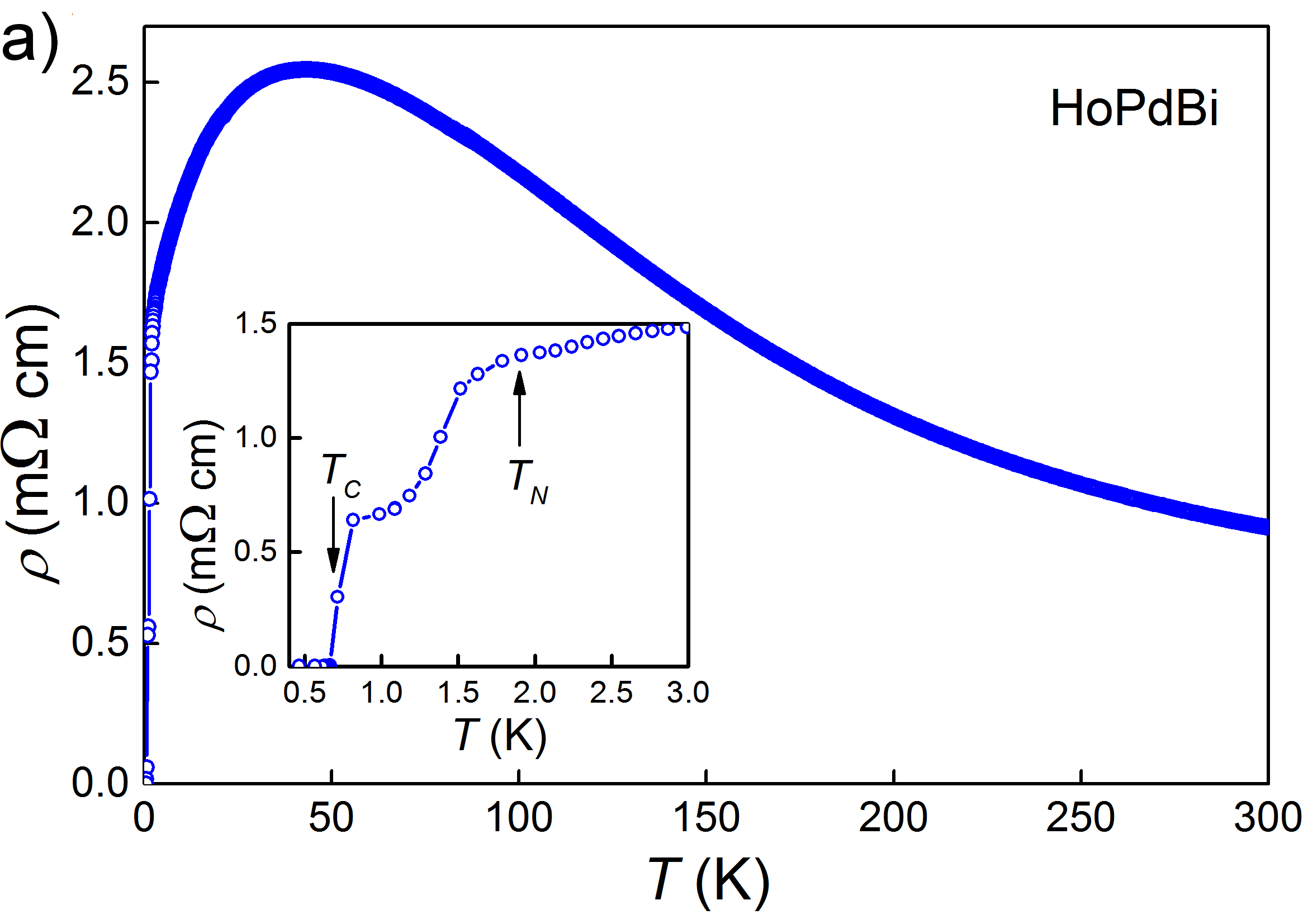}
\includegraphics[width=0.49\textwidth]{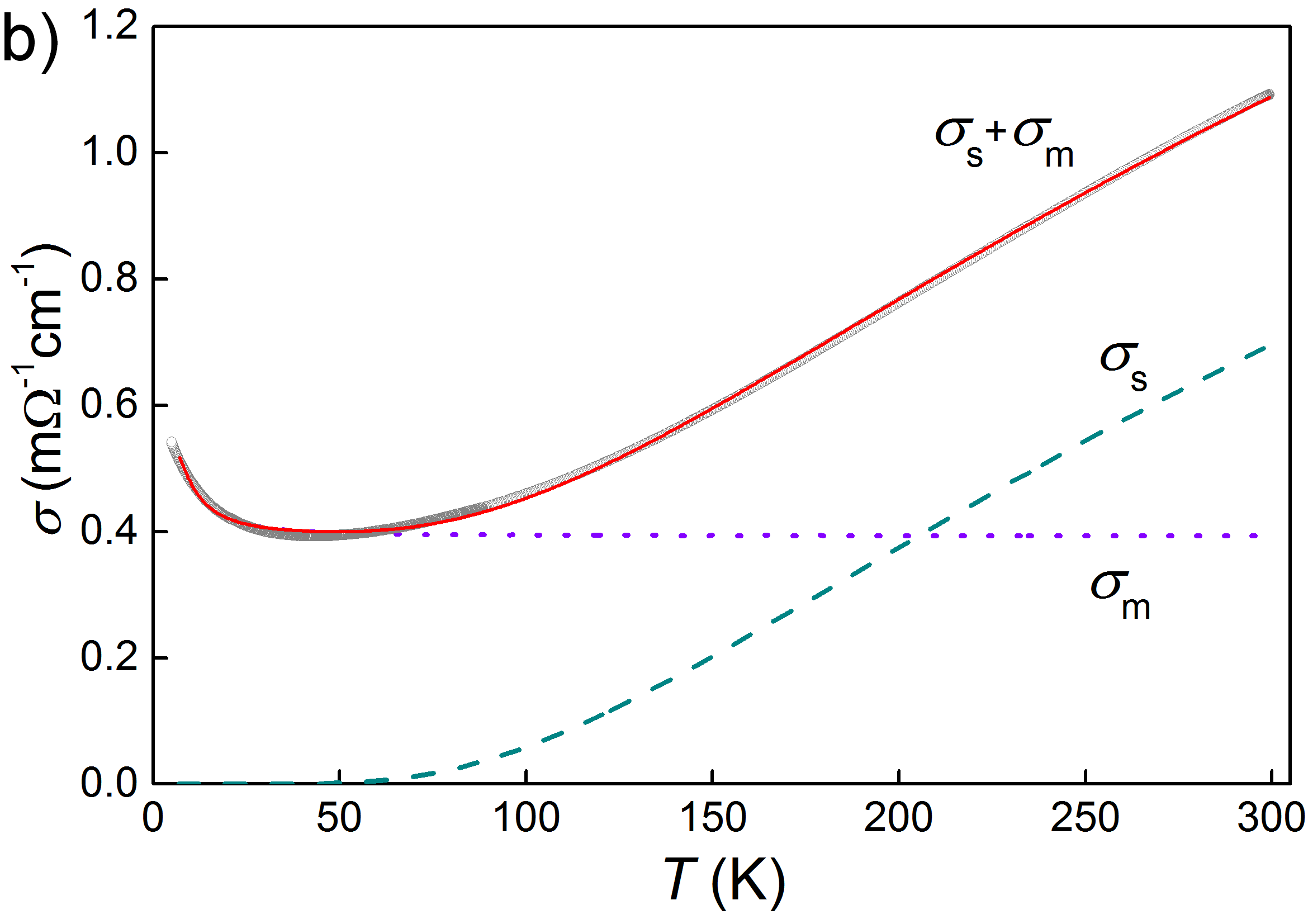}
\caption{a) Temperature variation of electrical resistivity, inset shows low temperature region with magnetic ordering and superconducting transitions indicated by arrows. b) Conductivity versus temperature plot showing fitted sum (solid red line) of semiconducting ($\sigma_{\rm s}$ - dashed green  line) and metallic ($\sigma_{\rm m}$ - dotted purple line) components described in text.
\label{rho}}
\end{figure}

Temperature dependence of the electrical resistivity of HoPdBi is depicted in Fig.~\ref{rho}a. At room temperature $\rho=1.0\,$m$\Omega$cm and increases with decreasing $T$ until it reaches 2.5\,m$\Omega$cm value at $T$ = 40\,K, from that point the sample demonstrates metallic-like behavior that is the resistivity decreases with temperature decreasing. The shape of $\rho(T)$ curve for HoPdBi is similar to those for other half-Heusler compounds containing rare earths.\cite{Gofryk2011, Pan2013a, Pavlosiuk2015, Pavlosiuk2015a, Nikitin2015, Nakajima2015}
Following the method we have applied before for LuPdBi, a nonmagnetic analogue of HoPdBi, we fitted the conductivity plotted versus temperature with a sum of two functions, $\sigma_{\rm m}$ and $\sigma_{\rm s}$, corresponding to two independent channels of charge transport: metallic- and semiconducting-like, respectively.\cite{Pavlosiuk2015} We defined the former term as
$\sigma_{\rm m} (T) =(\rho _0 + bT^2 +cT)^{-1}$, where $\rho _0$ is the residual resistivity due to scattering on structural defects, $bT^2$ represents electron-electron scattering processes, while $cT$ accounts for scattering on phonons. Contribution of the semiconducting channel was written as $\sigma_{\rm s}(T)=\sigma_0\exp(E_g/2k_{\rm B}T)$, where $E_g$ is an energy gap between valence and conduction bands. Considering the effect of strong crystal electric field (CEF) for holmium, we added to both $\sigma_{\rm m}$ and $\sigma_{\rm s}$ additional contributions, proportional (in terms of the resistivity) to $cosh(\Delta/T)^{-2}$. Such a phenomenological description of the resistivity resulting from scattering on CEF levels of 4$f$-electrons has been proposed in Ref.~\onlinecite{rhoCEF1963} and is in perfect agreement with quantum-mechanical calculations of Hessel Andersen et al.  \cite{rhoCEF1980}  Finally,
$\sigma_{\rm m}(T)=(\rho _0 + bT^2 +cT + p_{\rm m}cosh(\Delta/T)^{-2})^{-1}$ and
$\sigma_{\rm s}(T)=(\sigma_0^{-1}\exp(-E_g/2k_{\rm B}T)+p_{\rm s} cosh(\Delta/T)^{-2})^{-1}$ ($p_{\rm m}$ and $p_{\rm s}$ represent contributions of CEF effect to $\sigma_{\rm m}$ and $\sigma_{\rm s}$, respectively).
Fitting $\sigma(T)=\sigma_{\rm s}(T)+\sigma_{\rm m}(T)$ to the experimental data of HoPdBi in the temperature interval from 5 K to 300 K, yielded the parameters:  $\sigma_0 = 2.4\,{\rm m\Omega^{-1}cm^{-1}}$, $E_g=$\,64\,meV,
$\rho_0=1.77\,{\rm m\Omega cm}$ and $\Delta=10$\,K. Parameters $b$, $c$ and $p_{\rm s}$ given by the fit were all virtually zero, meaning that CEF effect brings significant contribution only to $\sigma_{\rm m}$ and that electron-electron and electron-phonon scattering are both negligibly small compared to magnetic scattering on CEF levels. The value of $E_g$ is similar to the values reported for other $RE$PdBi phases.\cite{Gofryk2005,Kaczorowski2005,Gofryk2011, Pan2013a,Pavlosiuk2015}
 
Position of broad maximum in the $\rho(T)$ dependence ($\approx40\,$K) is almost the same as for HoPdBi single crystals studied by Nakajima et al.\cite{Nakajima2015} but differs from $\approx75\,$K reported by Nikitin et al.\cite{Nikitin2015} In wide temperature range, from 300\,K to 2\,K, behavior of the resistivity of our single crystals is also more similar to that reported in Ref.~\onlinecite{Nakajima2015} than to that of Ref.~\onlinecite{Nikitin2015}. Values of $\rho$ of our single crystals are the largest amongst hitherto reported for HoPdBi,\cite{Gofryk2011,Nakajima2015, Nikitin2015} which seems to indicate that their semiconducting bulk has the largest gap among all studied specimens, which can be attributed to the smallest content of parasitic bulk metallic phases.

When our sample was cooled below 2\,K a sharp decline of its resistivity was observed - within 1\,K interval it dropped by half (see the inset of Fig.~\ref{rho}a). This behavior seems to be a consequence of antiferromagnetic ordering that sets in at $T\approx 1.95\,$K.\cite{Nakajima2015, Nikitin2015} We observed also a narrow plateau of $\rho(T)$ between 1.2 and 0.8\,K - in contrast to other reports,\cite{Nakajima2015, Nikitin2015} where decrease of $\rho(T)$ towards zero (starting at $T \approx1.4\,$K and at $T \approx2.3\,$K, respectively) was continuous and smooth.
Upon further cooling, the resistivity of our sample continued to decrease and at $T=0.67$\,K reached zero, indicating the emergence of superconductivity (see the inset of Fig.~\ref{rho}a). Nakajima et al. in Ref.~\onlinecite{Nakajima2015} reported on zero resistivity at $\approx0.7\,$K, very close to our result, but superconducting transition they described was rather wide, similar to the transition reported in Ref.~\onlinecite{Nikitin2015}, with the important difference that in the latter study $\rho$ attained the zero value already at $T\approx1.6\,$K. 
\subsection*{Magnetic properties}
\begin{figure}[b]
\includegraphics[width=0.49\textwidth]{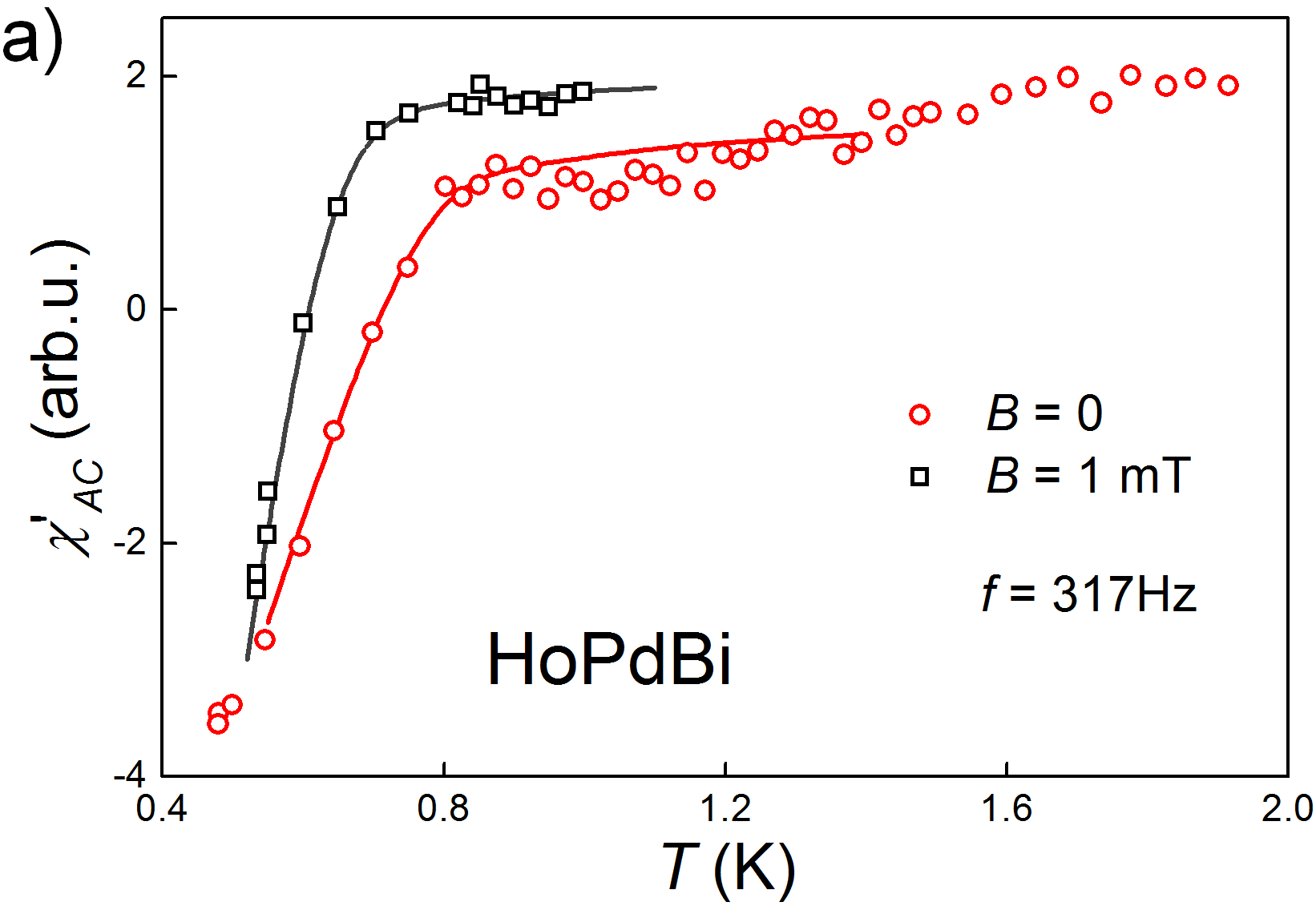}
\includegraphics[width=0.49\textwidth]{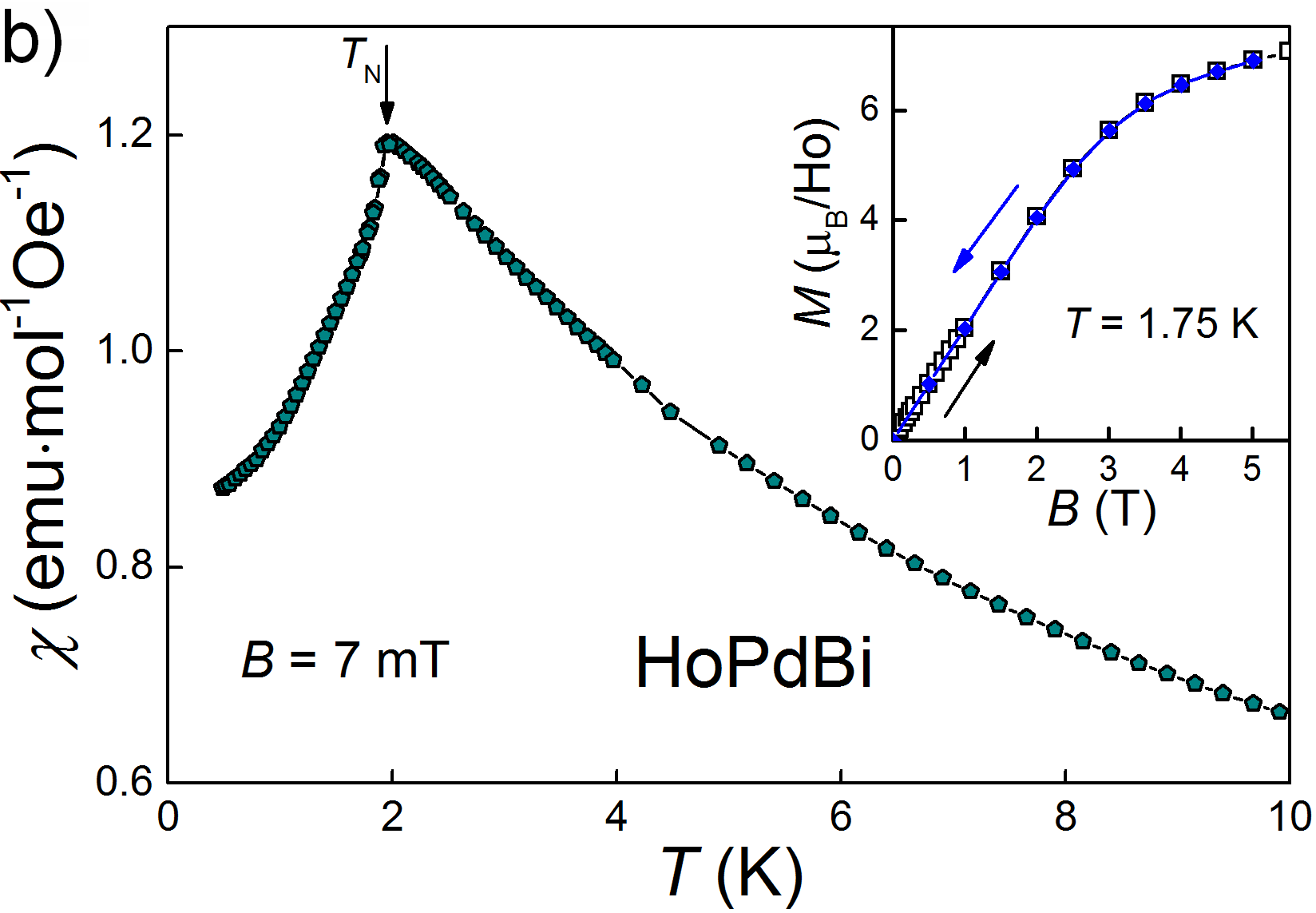}
\caption{a) Temperature dependence of AC magnetic susceptibility, solid lines are guides for eye. b) DC magnetic susceptibility versus temperature measured at magnetic field $B=7$\,mT, arrow indicates the N\'eel temperature. Inset: DC magnetization versus magnetic field recorded at $T=1.75$\,K, in increasing (open black squares) and decreasing (full blue diamonds) field.}
\label{CHI}
\end{figure}

Behavior of the low temperature dependence of real part of AC magnetic susceptibility, $\chi'_{AC}(T)$, (Fig.~\ref{CHI}a) affirms superconducting phase transition in HoPdBi,  as a keen reduction of $\chi'_{AC}$ appears below $T$ = 0.8\,K. Application of  the magnetic field, $B$, of 1\,mT shifts $T_c$ down by 0.1\,K. In parallel, temperature dependence of DC magnetic susceptibility, taken in 7\,mT (Fig.~\ref{CHI}b), down to 0.5\,K, does not show any anomaly that might correspond to $T_c$. There is a sharp peak at $T\approx 1.95\,$K, indicating antiferromagnetic ordering, in a perfect agreement with data reported previously. \cite{Nikitin2015, Gofryk2011, Nakajima2015}

The inset of Fig~\ref{CHI}b shows a field dependence of the magnetization, $M$, measured at $T = 1.75\,$K, in increasing and decreasing magnetic field. At low magnetic fields $M$ grows linearly, but at stronger fields $M(B)$ undergoes a flexure (around $B\approx3\,$T) and tends to saturation. The magnetic moment measured at $B=5$\,T equals $6.9\,\mu_B$ per holmium atom. This value is significantly smaller than theoretical one for free Ho$^{3+}$ ion ($g\cdot J = 5/4\cdot 8= 10\,\mu_B$).  Such difference is most likely due to splitting of ground multiplet in a cubic crystal field potential, which also has an impact on the electrical resistivity (see above) and the specific heat behavior (described below). The $M(B)$ isotherm does not show any hysteresis nor remanence. There is no visible sign of metamagnetic transition.
\subsection*{Specific heat}
\begin{figure}[h]
\includegraphics[width=0.49\textwidth]{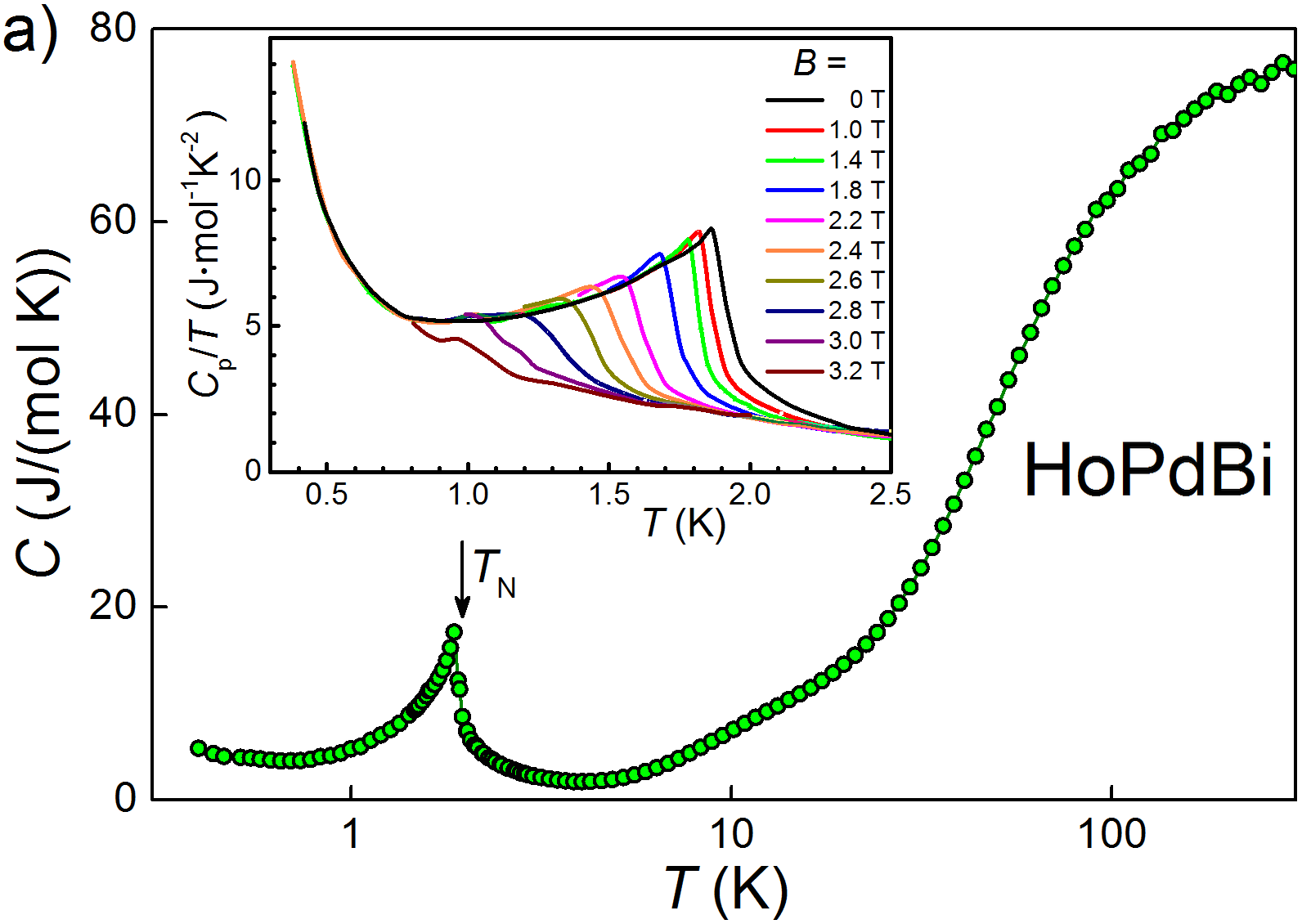}
\includegraphics[width=0.49\textwidth]{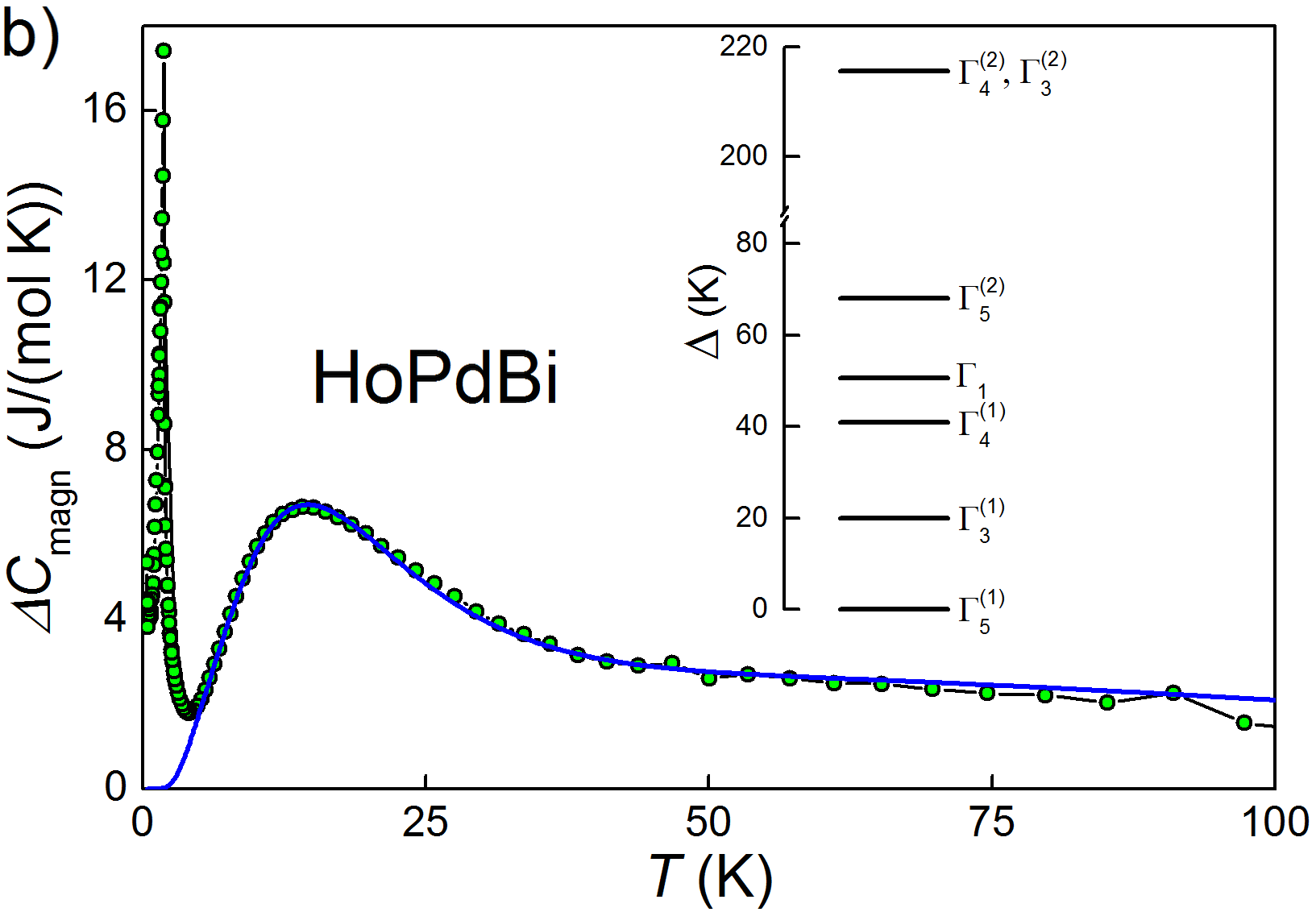}
\caption{a) Temperature variation of specific heat; inset: data collected at low temperatures and in applied magnetic fields. b) Magnetic component of specific heat. CEF contribution, fitted with Schottky function (Eq.\,\ref{Schot_eq}), is shown as a blue line. Inset shows the sequence of CEF levels resulting from that fit.}
\label{HC}
\end{figure}

The temperature dependence of the specific heat, $C(T)$, of HoPdBi in zero magnetic field is displayed in Fig.~\ref{HC}a. It shows a $\lambda$-shaped anomaly near $T_{\rm N}$ = 1.9\,K that can be ascribed to the antiferromagnetic phase transition, yet no singularity could be discerned at the critical temperature $T_{\rm c}$ = 0.7\,K. In the inset to Fig.~\ref{HC}a we present the effect of applied magnetic field on the AFM phase transition. Magnetic field of 1\,T has little influence on $T_{\rm N}$ but the 2\,T field shifts down the transition temperature to $\approx1.6$\,K. At 3\,T a small hump on the $C(T)$ curve is still noticeable at about $T=1$\,K, most likely reflecting the AFM phase transition. The $T_{\rm N}(B)$ behavior observed here agrees perfectly with phase diagram presented in Ref.~\onlinecite{Nikitin2015}. 

In addition to the distinct anomaly at $T_{\rm N}$, the $C(T)$ curve exhibits two other features, namely an upturn below about 0.6\,K, and a hump near 15\,K. The former effect most likely results from nuclear hyperfine interactions, which are very strong in holmium. Similar nuclear Schottky anomaly was observed in pure holmium\cite{Krusius1969} and in Ho-bearing materials, e.g. borocarbides.\cite{Rapp1999} For HoPdBi the low temperature upturn in $C(T)$ was  reported also by Nakajima et al.\cite{Nakajima2015} It is worth noting that our results clearly show that the upturn position is insensitive to magnetic field of up to 3\,T, indicating that the nuclear magnetic moment of holmium is very stable.

In turn, the shoulder of $C(T)$ around 15\,K seems to be the Schottky anomaly arising from the splitting of the ground-state multiplet of Ho$^{3+}$ due to crystalline electric field. Very similar hump in the specific heat of HoPdBi has been observed by Nikitin et al.~\cite{Nikitin2015} Assuming that the phonon contribution to the specific heat of HoPdBi is equal to that in the isostructural nonmagnetic bismuthide LuPdBi, the magnetic contribution to the specific heat of HoPdBi was calculated as $\Delta C_{\rm magn}=C^{\rm HoPdBi} - C^{\rm LuPdBi}$, where the data of LuPdBi were those presented in Ref.~\onlinecite{Pavlosiuk2015}. The resulting $\Delta C_{\rm magn}(T)$ data are shown in Fig.~\ref{HC}b. At temperatures above 6\,K they can be very well fitted with the Schottky formula\cite{Tari2003}:
\begin{equation}
\Delta C_{\rm magn}=R\,\frac{\Sigma_i g_i {\rm e}^{-\Delta_i/T} \Sigma_i g_i \Delta_i^2 {\rm e}^{-\Delta_i/T} - [\Sigma_i g_i \Delta_i {\rm e}^{-\Delta_i/T}]^2}{T^2 [\Sigma_i g_i {\rm e}^{-\Delta_i/T}]^2}
 \label{Schot_eq}
 \end{equation}
where $R$ is the gas constant, $\Delta_i$ is the energy of separate crystal field level and $g_i$ is its degeneracy. In the case of cubic crystal field potential, the ground multiplet $^5I_8$ of Ho$^{\rm 3+}$ ion splits into four magnetic triplets (two $\Gamma_5$ and two $\Gamma_4$), two non-magnetic doublets ($\Gamma_3$) and one non-magnetic singlet ($\Gamma_1$).\cite{Lea1962}
The CEF scheme derived for HoPdBi by fitting Eq.~\ref{Schot_eq} is shown in the inset to Fig.~\ref{HC}b.
The electronic ground state in HoPdBi is triplet $\Gamma_5^{(1)}$, the first excited state is a $\Gamma_3^{(1)}$ doublet, located at $\Delta_1$ = 22\,K, while the highest level at $\Delta_5$ = 216\,K is a pseudodegenerate state consisting of $\Gamma_3^{(2)}$ and $\Gamma_4^{(2)}$ states. Remarkably, a similar CEF splitting scheme was established before by means of inelastic neutron scattering for a closely related compound HoNiSb.\cite{Karla1999}
\subsection*{Neutron diffraction}
\begin{figure}[h]
\includegraphics[width=8cm]{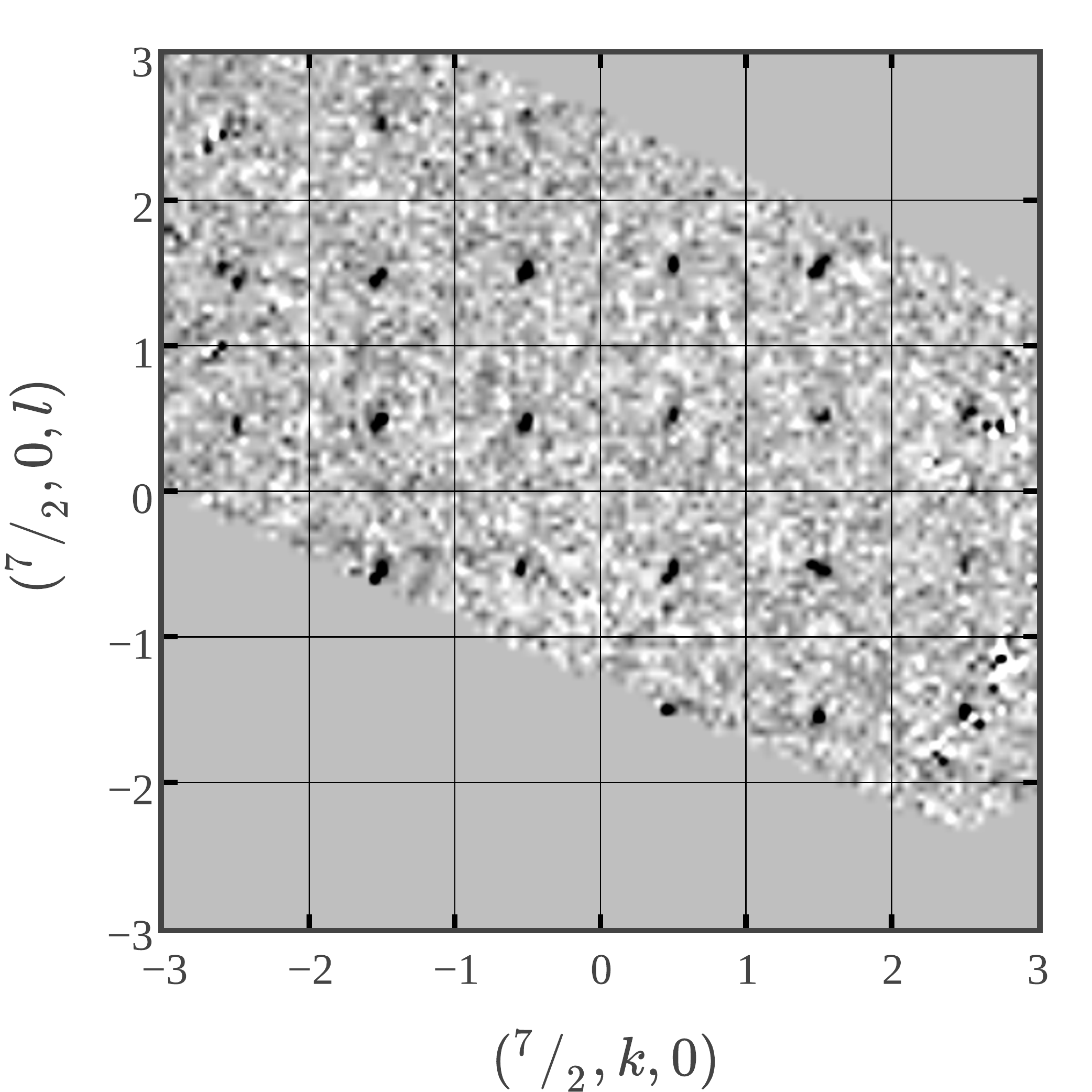}
\caption{The ($\rfrac{\,7}{2},\,k,\,l$) differential (magnetic) neutron diffraction map on HoPdBi obtained by subtraction of the intensities collected at $T=6$\,K from those collected at $T=1.6$\,K. 
\label{neut}}
\end{figure}

Neutron diffraction on our HoPdBi single crystal confirms the antiferromagnetic order below 1.9\,K. We performed measurements below and above $T_{\rm N}$ and observed additional reflections on diffraction pattern at $T=1.6$\,K which were not present at temperatures higher than $T_{\rm N}$. As an example in Fig.~\ref{neut} we show differential intensity map resulting from subtraction of experimental data collected at $T=6$\,K from those collected at $T=1.6$\,K. Positions of obtained reflections clearly indicate the propagation vector $(\rfrac{\,1}{2},\rfrac{1}{2},\rfrac{1}{2})$ prerequisite for AFTI state proposed by the theory of Mong et al.\cite{Mong2010}. Very recently such a structure has been suggested, based on neutron diffraction on polycrystalline sample of HoPdBi, but without any data presented.\cite{Nakajima2015}
\subsection*{Magnetoresistance}
\begin{figure}[h]
\includegraphics[height=6cm]{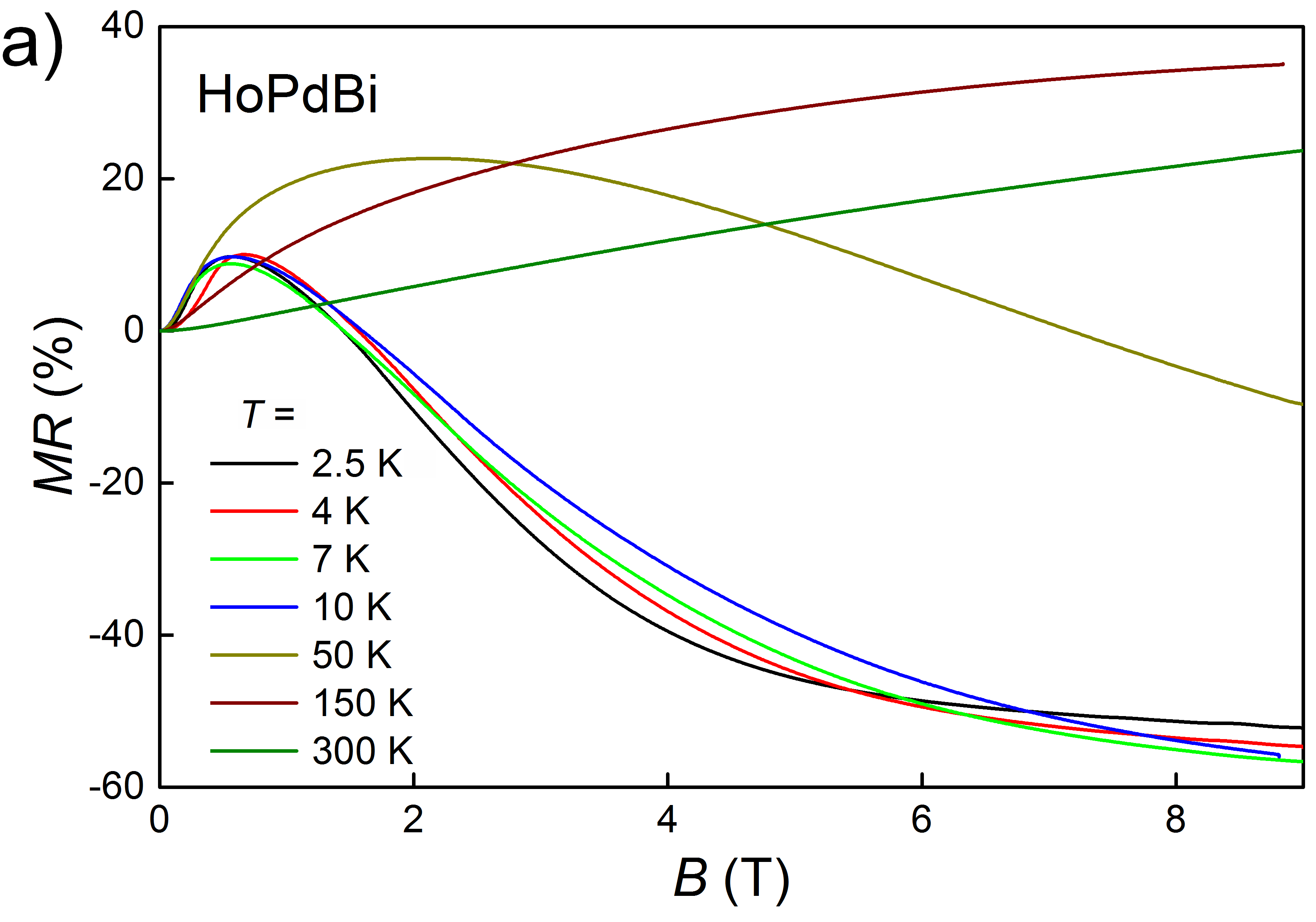}
\includegraphics[height=6cm]{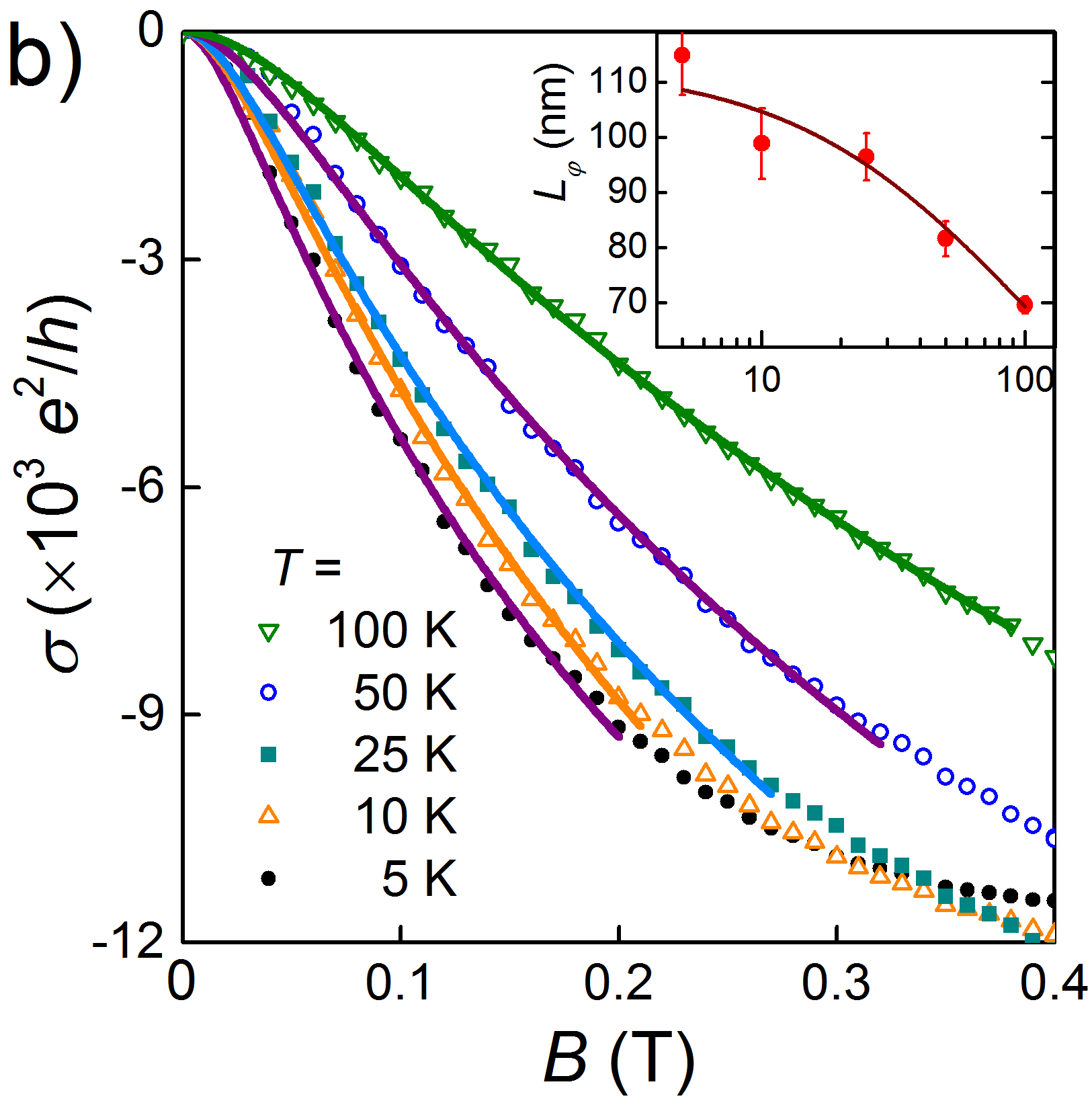}
\caption{a) Magnetoresistance versus applied magnetic field at temperatures between $T=2.5$\,K and 300\,K. b) Low-field dependence of magnetoconductivity at 5, 10, 25, 50 and 100\,K. Solid lines show the fits with the Hikami-Larkin-Nagaoka function (Eq.\ref{WALeq}). Inset: temperature dependence of phase coherence lenght obtained from these fits.
\label{MR}}
\end{figure}

The transverse magnetoresistance, $M\!R=[\rho(B)-\rho(0)]/\rho(0)$, measured at several temperatures in the paramagnetic state is plotted versus magnetic field in Fig.~\ref{MR}a. The curves have unusual shapes, quite different from those predicted by classical theory of $M\!R$. At temperatures from 2.5 to 10\,K $M\!R(B)$ curves differ very little, $M\!R$ quickly increases with increasing  magnetic field and reaches a maximum of 9\,\% at $\approx\!0.5\,$T. Such sharp increments of resistivity at low magnetic field may originate from a weak antilocalization (WAL) effect, which appears when probability of electrons to become localized is reduced as a result of destructive interference of electron wave function. Previously WAL was observed in different topological states of matter, in topological insulators\cite{Taskin2012} and Weyl semimetals\cite {Yang2015}. The Hikami-Larkin-Nagaoka (HLN) theory\cite{Hikami1980} is considered the most suitable for describing systems in which 2D WAL appears. We also applied that theory to our system, by fitting the sheet conductivity with the HLN formula:
\begin{equation}
\Delta\sigma(B)=\frac{-\alpha e^2}{2\pi^2 \hbar}\,\Big[\psi\Big(\frac{1}{2}+\frac{\hbar}{4eL_\varphi^2B}\Big)-ln\Big(\frac{\hbar}{4eL_\varphi^2B}\Big)\Big],
\label{WALeq}
\end{equation}
where the prefactor $\alpha$ is accountable for the mechanism of localization, $L_\varphi$ is the phase coherence length (see Fig.~\ref{MR}b) and $\psi$ is digamma function. For $T=10$\,K we obtained $L_\varphi=98$\,nm and $\alpha$ of the order of $10^4$. Probably, such big value of $\alpha$ (compared to 1/2 expected for a 2D system) is brought about by bulk and side wall conductivity.\cite{Sacksteder2014, Xu2014a, Pavlosiuk2015} Prefactor $\alpha$ is almost independent of temperature, that  indicates robustness of surface states. In turn, $L_\varphi$ decreases monotonously with increasing temperature from 115\,nm at 2.5\,K to 70\,nm at 100\,K (see inset to Fig.~\ref{MR}b),
this behavior of $L_\varphi(T)$ is similar to that observed for LuPdBi.\cite{Pavlosiuk2015} At higher magnetic fields $M\!R$ decreases monotonously, passing through zero at  $\approx\!1.7\,$T, and attaining negative values of $\approx\!-50\,$\% in fields above 6\,T, where its decreasing becomes significantly slower but continues without saturation with the magnetic fields increasing up to 9\,T.

%
At 50\,K the initial slope of $M\!R(B)$ is similar to that at 10\,K, but $M\!R$ continues increasing in fields above 0.5\,T, to reach a maximum of 22\,\% at $B=2\,$T. Subsequent decreasing of $M\!R$ leads to zero value at $B=7\,$T and $-10\,\%$ at 9\,T.
From $T=150$\,K up to room temperature $M\!R$ is positive. At $T=300\,$K and magnetic field of 9\,T, $M\!R=24\,\%.$
Overall behavior of $M\!R$ is correlated with behavior of the resistivity, namely at temperatures below the broad maximum in $\rho(T)$ $M\!R$ is dominated by negative component, at 50K the positive component starts to overcome the negative one, and at higher temperatures $M\!R$ becomes positive.

Such unusual and complicated behavior of transverse $M\!R$ has never been observed before for rare earth palladium bismuthides. Negative $M\!R$ has been reported for polycrystalline samples of other antiferromagnetic $RE$PdBi ($RE=$Nd, Gd, Dy, Er) but, unlike our data, it was negative in the whole range of magnetic field.\cite{Riedemann1996} Gofryk et al. observed qualitatively similar but much less pronounced shape of $M\!R$ isotherms in ErPdSb.\cite{Gofryk2007} On the other hand, very similar $M\!R(B)$ curves have been reported for a heavy-fermion compound YbPtBi.\cite{Mun2013} Although the low-field maximum was not present, already at field of 4\,T the transverse $M\!R\approx\!-50\%$ was attained. 
Furthermore, very similar transverse $M\!R$ curves have been observed in uniaxially compressed single crystals of HgTe (with $M\!R\approx\!-40\%$)\cite{Takita1974} and in single crystalline Bi$_2$Se$_3$ under hydrostatic pressure ($M\!R\approx\!-10\%$).\cite{Hamlin2012} Authors of Ref.~\onlinecite{Takita1974} have attributed such $M\!R(B)$ behavior to the increase of the electron population of the ground Landau level with increasing magnetic field resulting in closing a small (2--3\,meV) gap  between conduction and valence bands. That does not seem adequate for HoPdBi with the estimated gap of 64\,meV (see the discussion of $\rho(T)$ above). Recently, a theory of transversal $M\!R$ in Weyl semimetals was proposed,\cite{Klier2015} showing that the presence of 'short-range' point-like (non-Coulomb) impurities may cause $M\!R$ to reach a maximum and subsequently decrease with magnetic field increasing. However, position of $M\!R(B)$ maximum is scaled by $T^2$ in that theory, whereas in case of HoPdBi maximum hardly moves with $2.5\leq T\leq 10$.

A decrease of $M\!R$ upon increasing magnetic field might also be attributed to antiferromagnetic fluctuations but it is rather unlikely to observe such effect even at $T=50$\,K -- significantly higher than $T_{\rm N}$.
The $M\!R(B)$ behavior nearly identical to that of HoPdBi has been reported for a similar half-Heusler phase -- HoNiSb (semiconductor with a gap of 61 meV).\cite{Karla1998}
Authors of that article have attributed strong negative $M\!R$ at higher fields to the reduction of spin-disorder scattering due to the alignment of moments. Such model seems reasonable also for HoPdBi and, together with WAL effect causing positive $M\!R$ at low fields, would explain complicated $M\!R(B)$ curve in the whole range of fields covered by our measurements. We fitted the de Gennes-Friedel function described in Ref.~\onlinecite{Karla1998} to our $\rho(B)$ data and results of these fits are presented in Supplementary Material.
\subsection*{Shubnikov-de Haas oscillations}
\begin{figure}[t]
\includegraphics[width=12cm]{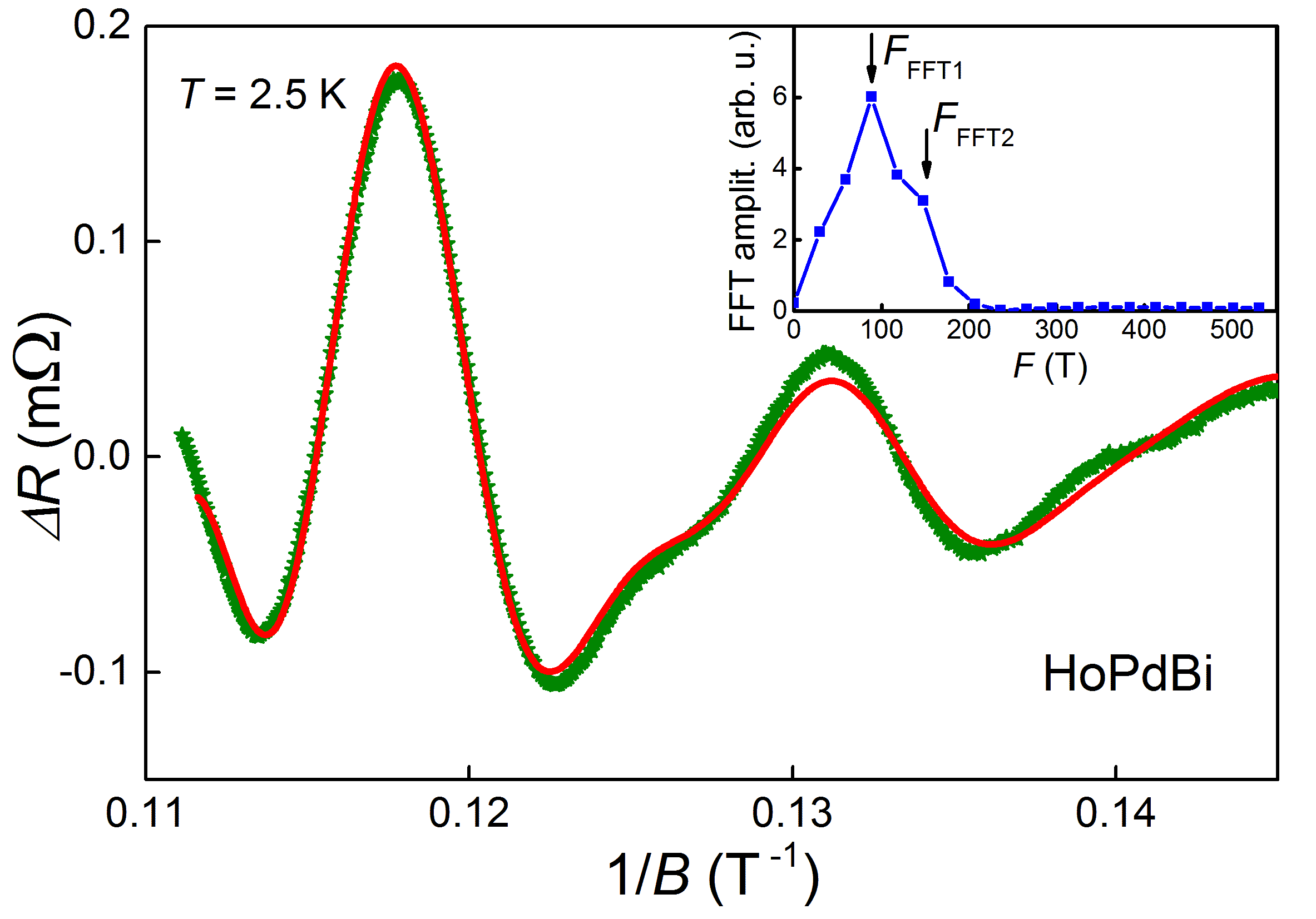}
\caption{Oscillatory component of the resistance at $T=2.5$\,K. Solid red line shows the fit with the modified Lifshitz-Kosevich expression (see text). Inset: the fast Fourier transform spectra comprising a broad peak with a shoulder. Black arrows indicate two oscillatory frequencies.
\label{SdH}}
\end{figure}

Oscillations of the resistance in changing magnetic fields, i.e. Shubnikov-de Haas (SdH) effect is a powerful experimental tool for investigations of topological insulators\cite{Ando2013} and Dirac semimetals.\cite{Potter2014a} In the latter case surface Fermi arcs connected by trajectories crossing bulk of the sample may form closed orbits  resulting in quantum oscillations. In the case of HoPdBi, SdH oscillations could be traced down to a field $\sim$6\,T and up to temperature of 7\,K. Clear periodic fluctuation of the resistivity as a function of 1/$B$ is observed after background removal (Fig.~\ref{SdH}). Fast Fourier transform (FFT) results  presented in the inset of Fig.~\ref{SdH}, show a broad peak with maximum at $F_{\rm FFT1}$ = 87\,T and a shoulder near $F_{\rm FFT2}$ = 147\,T. We fitted experimental data with a superposition of two Lifshitz-Kosevich expressions\cite{Shoenberg1984} (represented by red line in Fig.~\ref{SdH}), each corresponding to a single cross-section of Fermi surface. The obtained fit parameters and the resulting physical quantities for $T$ = 2.5\,K are listed in Table~\ref{PSdH}. The frequencies estimated from the fit, $F_1$ = 78.5\,T and $F_2$ = 143.9\,T, coincide  well with those extracted from the FFT spectrum. $F_1$ is in a perfect agreement with the frequency value reported by Nikitin et al.\cite{Nikitin2015}, however, they did not observe the second frequency. It is worth noting that the amplitude of SdH oscillations observed by us at $B\approx9$\,T and at $T=2.5$\,K is one order of magnitude larger than the amplitude of oscillations found in Ref.~\onlinecite{Nikitin2015} at much lower temperatures ($\leq0.6\,$K).

For both oscillatory components a phase factor, $\gamma$, is not zero but 0.65 and 0.38 for $F_1$ and $F_2$, respectively. Non-zero phase factor indicates that the system may hosts Dirac fermions. Moreover, for 3D Dirac-like carriers $\gamma = \rfrac{1}{2}\pm \rfrac{1}{8}$ (+ sign for holes and $-$ sign for electron carriers),\cite{Lukyanchuk2004} unlike 2D Dirac fermions having $\gamma=\rfrac{1}{2}$. Hall efect measurements indicated that in HoPdBi current carriers are holes,\cite{Nikitin2015} so in this case $\gamma$ should be $\rfrac{1}{2}+\rfrac{1}{8}= 0.625$. The obtained value $\gamma = 0.65$ is very close to theoretical one, thereby we may suppose that HoPdBi hosts 3D Dirac fermions. Fermi wave vectors, $k_{\rm F}$, calculated from the Onsager relation, $F= (\hbar/2e)k^2_{\rm F}$, are $4.8\times10^6$ and $6.6\times10^6\,$cm$^{-1}$ for $F_1$ and $F_2$, respectively. They correspond to the carrier density $n_{\rm 3D}=k^3_{\rm F}/3\pi^2$ equal to $3.7\times10^{18}$ and $9.8\times10^{18}\,$cm$^{-3}$, respectively. Sum of these two densities ($1.35\times10^{19}\,$cm$^{-3}$) is practically identical to the carrier concentration $n_{\rm H}\approx1.3\times 10^{19}\,$cm$^{-3}$ determined by Hall effect measurements.\cite{Nikitin2015}
If one assumes that the SdH oscillations originate from spin-nondegenerate surface states, 2D carrier densities following from $F_1$ and $F_2$ are both of the order of $10^{12}\,$cm$^{-2}$.
Another parameter of interest is cyclotron mass, $m_c$, which can be obtained from the temperature dependence of amplitudes of SdH oscillations. Carriers that enclose smaller orbits have $m_c=0.44\,m_{\rm e}$ ($m_{\rm e}$ is the free electron mass), and this value is consistent with that reported for HoPdBi by Nikitin et al.\cite{Nikitin2015}
Cyclotron mass corresponding to the second frequency is smaller and equals 0.28$\,m_e$. Both masses are similar to those determined for other rare-earth-containing half-Heusler phases.\cite{Pavlosiuk2015, Pavlosiuk2015a, Butch2011a, Wang2013, Wosnitza2006a} Once $m_c$ is known, the so-called Dingle plot can be used to determine the Dingle temperature, $T_D$, from which the lifetime $\tau = \hbar/2\pi k_{\rm B}T_D$ and the carrier mobility $\mu = e\tau/m_c$ can be calculated. The values derived for HoPdBi, together with the Fermi velocity $v_{\rm F}= \hbar k_{\rm F}/m_c$, the Fermi energy $E_{\rm F}= m_cv_{\rm F}^2$ and the mean-free path $l=v_{\rm F}\tau$ are collected in Table~\ref{PSdH}.
\begin{table}[h]
\caption{Parameters derived from the analysis of SdH oscillations.}
{\footnotesize 
\begin{tabular}{ccccccccccc}
\\\hline\hline
$F_{i(=1,2)}$& $n_{3D}$& $m_c$& $k_F$& $T_D$& $\tau$& $V_F$& $E_F$& $l$& $\mu$\\
(T) & (cm$^{-3}$) & ($m_{\rm e}$)&(cm$^{-1}$)&(K)&(s)&(m/s)& (meV)& (nm)&  (cm$^2$V$^{-1}$s$^{-1}$)\\\hline
78.5 ~~& $3.7\times 10^{18}$ ~~& 0.44 ~~& $4.8\times 10^6$ ~~& 9.7 ~~& $1.25\times 10^{-13}$ ~~& $1.26\times 10^5$ ~~& 39.9 ~~& 16 ~~& 501\\
143.9 ~~& $9.8\times 10^{18}$ ~~& 0.28 ~~& $6.6\times 10^6$ ~~& 28.2 ~~& $4.31\times 10^{-14}$ ~~& $2.73\times 10^{5}$ ~~& 119 ~~& 12 ~~& 271\\
\hline\hline
\end{tabular}}\label{PSdH}\end{table}

The observation of SdH oscillations with two different frequencies hints at rather complex Fermi surface. The complexity may originate from the topologically semimetallic nature of HoPdBi. Any Dirac semimetal can be viewed as two copies of Weyl semimetals with opposite chiralities, and thus two sets of surface arcs. When magnetic field is weak, these two Fermi arcs are bound together through the bulk and oscillations with a single period arise. With magnetic field increasing, link between two surface arcs are broken and separate orbits on each surface appear, generating quantum oscillations with different period. Crossover from one set of orbits to another one may occur as a reentrant behavior of SdH oscillations or interference between two oscillations.\cite{Potter2014a} To check whether this concept works for HoPdBi, SdH measurements in high magnetic fields are necessary.
\section*{Conclusions}
Our examination of electronic transport, magnetic and thermal properties of flux-grown HoPdBi single crystals clearly shows the coexistence of antiferromagnetic and superconducting phases. We confirmed that HoPdBi orders antiferromagnetically at $T_{\rm N}=1.9$\,K and that the magnetic structure is characterized by the ($\rfrac{\,1}{2},\rfrac{1}{2},\rfrac{1}{2}$) propagation vector. Further magnetotransport studies at temperatures below $T_{\rm N}$ in comparison with our data collected at $T\geq2.5$\,K would be necessary to reveal how the magnetic ordering influences topological properties of HoPdBi. 

Our and previously published studies demonstrate very similar, sample independent, antiferromagnetic characteristics of HoPdBi.\cite{Nakajima2015,Nikitin2015} Our extended specific heat analysis shows that two features, one near 15\,K and another below 0.6\,K, which can be attributed to CEF and nuclear Schottky effect, respectively.

Most remarkably, the temperature dependence of the specific heat does not reveal any anomaly near $T_{\rm c}=0.7$\,K, the onset of SC state, clearly indicated by both the electrical resistivity and the AC magnetic susceptibility results. Lack of specific heat anomaly at $T_{\rm c}$ may be a consequence of the surface nature of superconductivity, intrinsic or induced by very weak magnetic fields (even by residual fields of superconducting magnets, being of order of 0.1\,mT). Discrepancies in results of numerous studies of SC in half-Heusler phases (not only HoPdBi, but also ErPdBi, LuPdBi and others)\cite{Pan2013a,Pavlosiuk2015a,Xu2014a,Pavlosiuk2015} could be explained if the superconducting phase was easily destroyed in the bulk by such weak fields, but survived on the surface, possibly due to topological nature of surface conducting channel. In order to prove this conjecture, very careful measurements at meticulously controled low magnetic fields are necessary.
Electronic state of half-Heusler phases is very sensitive to small, undetectable by standard methods, deviations of stoichiometry, as indicated by diverse temperature dependencies of the resistivity in different samples of nominally the same compound - this should also significantly influence characteristics of superconducting state.

We observed  WAL effect causing positive $M\!R$ in weak fields and strong, negative $M\!R$ at higher fields, which seems to be due to the reduction of spin-disorder  scattering. SdH oscillations with two frequencies and non-zero Berry phase suggest a topologically non-trivial nature of HoPdBi. The latter conjecture, however, requires verification by direct probes such as ARPES measurements.
\section*{Methods}
\subsection*{Material preparation}
The single crystals were grown from Bi flux with the starting composition Ho:Pd:Bi = 1:1:15 (atomic ratio). First, a polycrystalline ingot of HoPdBi was prepared by arc-melting almost equiatomic composition (with small excess of bismuth to compensate for loses by evaporation). Then the ingot was crushed and mixed with Bi grains. The charge was put in an alumina crucible and sealed inside an evacuated quartz ampule. The ampule was heated slowly to 1100$\,^{\circ}$C and kept for 24 hours, than slowly cooled down to 900$\,^{\circ}$C at 1$\,^{\circ}$C/hour and kept for 12 hours. The same cooling rate was down to 550$\,^{\circ}$C and the quartz tube was taken out from the furnace at this temperature. The excess of Bi flux was removed by centrifugation. Grown single crystals had shapes of cubes with dimensions up to 3$\times$3$\times$3\,mm$^3$.
\subsection*{Material characterization}
Composition of the HoPdBi single crystals was studied on a FEI scanning electron microscope (SEM) equipped with an EDAX Genesis XM4 energy-disspersive spectrometer (EDS). Specimens were glued to a SEM stub using carbon tape. EDS spectrum and SEM image are presented in Supplementary Materials. The crystals were found homogeneous and free of foreign phases.  
Crystal structure was examined by x-ray powder diffraction carried out on powdered single crystals, using an X'pert Pro PANanalytical diffractometer with Cu-K$\alpha$ radiation. Obtained diffractogram is shown in Supplementary Materials. Lattice parameter 6.613\,\AA\,  and space group $F$\={4}3{\it m} were determined, in a perfect agreement with previously reported data.\cite{Marazza1980}
\subsection*{Physical measurements}
DC magnetization and AC magnetic susceptibility measurements were carried out in the temperature interval 0.5--6\,K and in a wide range of magnetic fields using a  MPMS-XL SQUID magnetometer (Quantum Design) equipped with $^3$He refrigerator (iHelium3).

For the electrical transport measurements, bar shaped specimens were cut from single crystals and then polished. A standard four-probe method was used for resistivity measurements. Electrical leads were made from 50\,$\mu$m thick silver wires attached to the sample with dimensions 0.1$\times$0.4$\times$1.1\,mm$^3$ by spot welding and mechanically strengthened with silver paste. Experiments were performed in the temperature range 0.5--300\,K and in the magnetic fields up to 9\,T using a PPMS platform (Quantum Design) with a $^3$He refrigerator.

The specific heat experiments were carried out by relaxation method on a 2.2\,mg single crystal, within the temperature range 0.4--300\,K, also employing the PPMS platform. 

For the neutron diffraction experiment a high-quality single crystal with an approximate volume of 27\,mm$^3$ was selected. Measurements were carried out on VIP, a  two-axis neutron diffractometer at the Laboratoire L{\'e}on Brillouin, CEA-Saclay,\cite{Gukasov2013} using a neutron wavelength of 0.84\,\AA. Data were collected at two temperatures: 1.6 and  6\,K.

\newpage
\noindent \begin{large}{\bf SUPPLEMENTARY MATERIAL} \end{large}\\\\
{\bf Antiferromagnetism and superconductivity in the half-Heusler semimetal HoPdBi}\\\\
O. Pavlosiuk, D. Kaczorowski, X. Fabreges, A. Gukasov and P. Wi\'sniewski*

\section*{Material characterization}
\subsection*{Energy-dispersive X-ray spectroscopy}
Examples of the EDS spectrum and the SEM image obtained for the single crystals of HoPdBi are presented in Fig.\,S1. 
The chemical composition derived from EDS analysis is Ho$_{31.3(3)}$Pd$_{35.8(4)}$Bi$_{32.9(4)}$, in a fairly good accord with the ideal equiatomic one. The crystals were found homogeneous and no foreign phase was detected.
\begin{figure}[h]
\includegraphics[height=7cm]{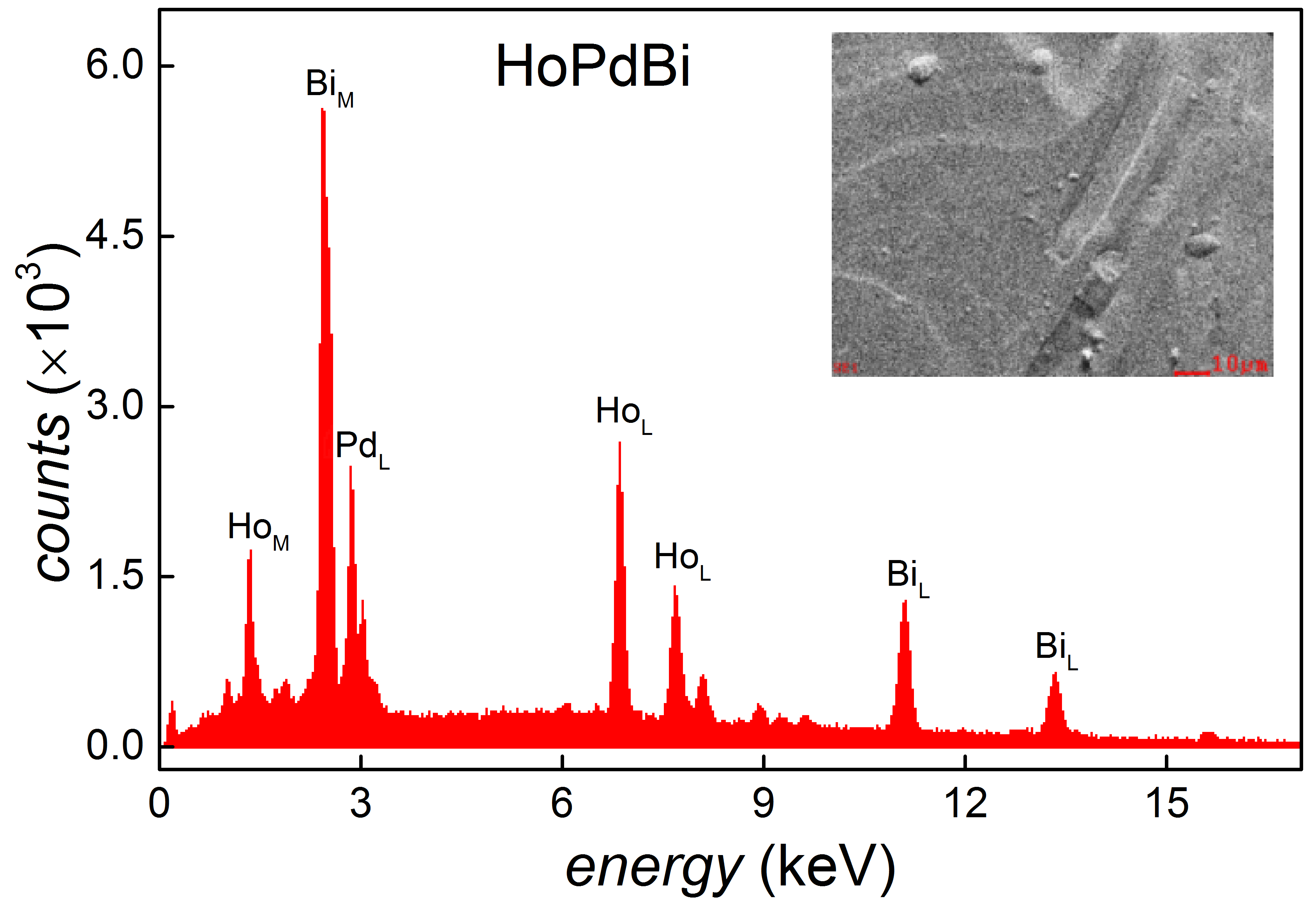}\\
{Figure S1. EDS spectrum and SEM image for single-crystalline HoPdBi.}
 \label{edx}
\end{figure}

\subsection*{X-ray diffraction}
The PXRD results (see Fig. S2) confirmed a single-phase character of the obtained single crystals of HoPdBi. The X-ray diffraction pattern can be fully indexed within the $F$\={4}3$m$ space group, characteristic of half-Heusler compounds, and yields the lattice parameter $a=\,6.613(2)\,$\AA. 
\begin{figure}[t]
\includegraphics[height=7cm]{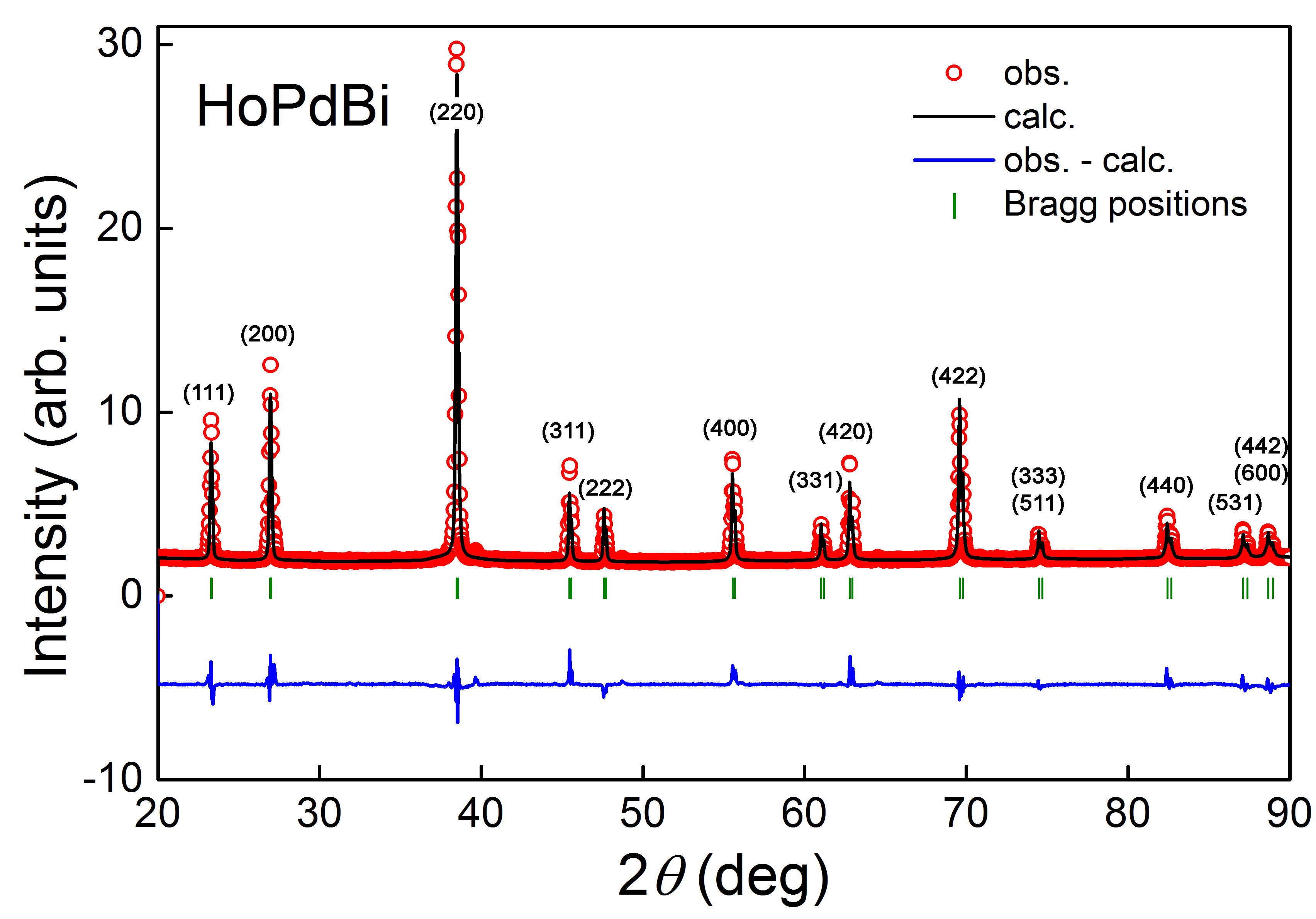}\\
{Figure S2. XRD diffractogram for powdered single crystals of HoPdBi.}
\label{XRDplot}
\end{figure}
This value is in perfect accord with the literature value $6.610(1)\,$\AA~determined for polycrystalline sample,\cite{Marazza1980} and the value $6.605\,$\AA~reported recently for for powdered single crystals.\cite{Nikitin2015} As can be inferred from Fig.~S2, the experimental PXRD pattern of our single-crystalline HoPdBi can be very well modeled with the MgAgAs-type crystal structure with the Ho atoms located at the crystallographic 4$a$ ($0, 0, 0$) sites, the Pd atoms occupying the 4$c$ ($1/4, 1/4, 1/4$) sites, and the Bi atoms placed at the 4$b$ ($1/2, 1/2, 1/2$) sites. 
\subsection*{Analysis of the negative magnetoresistance}
Following the analysis applied for HoNiSb in Ref.~\onlinecite{Karla1998} we fitted our $\rho(B)$ data with the de Gennes-Friedel function: 
$\rho(B)= \rho_0 [1-(\beta(CB/(T-\theta_{\rm CW})))^2]$, 
where $C=N\mu^2/k_{\rm B}$ is Curie constant and $\theta_{\rm CW}$ is paramagnetic Curie-Weiss temperature and $\beta$ denotes Brillouin function. 
Fitting data collected at 2.5\,K, using $\theta_{\rm CW}$ fixed at -9.4\,K determined in Ref.~\onlinecite{Nakajima2015}, gives a magnetic moment $\mu=5.96\mu_{\rm B}$ for Ho, close to 6.9\,$\mu_{\rm B}$ obtained by magnetization measurement but lower than  the theoretical value for a free Ho ion. This reduction may be justified by crystal field effects. That fit is shown in Fig.~S3, together with those performed for data collected at higher temperatures (yielding similar values of $\mu$). 
\newpage
\begin{figure}[h]
\includegraphics[height=7cm]{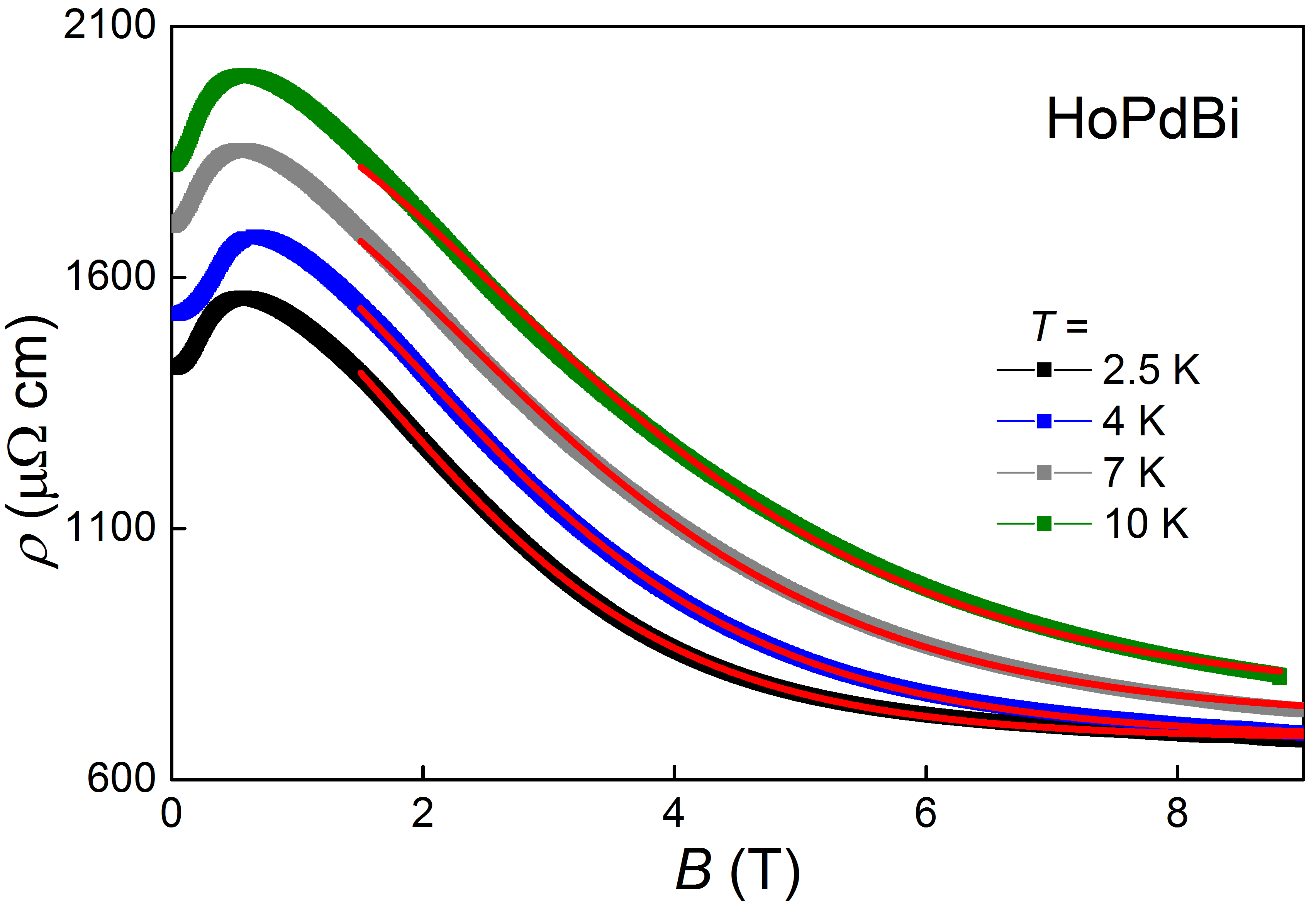}\\
{Figure S3. Field dependence of the resistivity of HoPdBi at different temperatures.\\ Red lines represent de Gennes-Friedel function fitted to experimental data.}
\label{negMR}
\end{figure}


\begin{thebibliography}{100}
\expandafter\ifx\csname url\endcsname\relax
  \def\url#1{\texttt{#1}}\fi
\expandafter\ifx\csname urlprefix\endcsname\relax\def\urlprefix{URL }\fi
\providecommand{\bibinfo}[2]{#2}
\providecommand{\eprint}[2][]{\url{#2}}

\bibitem{Pan2013a}
\bibinfo{author}{Pan, Y.} \emph{et~al.}
\newblock \bibinfo{title}{{Superconductivity and magnetic order in the
  noncentrosymmetric half-Heusler compound ErPdBi}}.
\newblock \emph{\bibinfo{journal}{Europhys. Lett.}}
  \textbf{\bibinfo{volume}{104}}, \bibinfo{pages}{27001}
  (\bibinfo{year}{2013}).
\bibinfo{doi}{DOI: 10.1209/0295-5075/104/27001}.

\bibitem{Goraus2013}
\bibinfo{author}{Goraus, J.}, \bibinfo{author}{\'{S}lebarski, A.} \&
  \bibinfo{author}{Fija{\l}kowski, M.}
\newblock \bibinfo{title}{{Experimental and theoretical study of CePdBi.}}
\newblock \emph{\bibinfo{journal}{J. Phys. Condens. Matter}}
  \textbf{\bibinfo{volume}{25}}, \bibinfo{pages}{176002}
  (\bibinfo{year}{2013}).
\bibinfo{doi}{DOI: 10.1088/0953-8984/25/17/176002}.

\bibitem{Nakajima2015}
\bibinfo{author}{Nakajima, Y.} \emph{et~al.}
\newblock \bibinfo{title}{{Topological RPdBi half-Heusler semimetals: A new
  family of noncentrosymmetric magnetic superconductors}}.
\newblock \emph{\bibinfo{journal}{Sci. Adv.}} \textbf{\bibinfo{volume}{1}},
  \bibinfo{pages}{e1500242} (\bibinfo{year}{2015}).
\bibinfo{doi}{DOI: 10.1126/sciadv.1500242}.

\bibitem{Nikitin2015}
\bibinfo{author}{Nikitin, A.~M.} \emph{et~al.}
\newblock \bibinfo{title}{{Magnetic and superconducting phase diagram of the
  half-Heusler topological semimetal HoPdBi}}.
\newblock \emph{\bibinfo{journal}{J. Phys. Condens. Matter}}
  \textbf{\bibinfo{volume}{27}}, \bibinfo{pages}{275701}
  (\bibinfo{year}{2015}).
\bibinfo{doi}{DOI: 10.1088/0953-8984/27/27/275701}.

\bibitem{Chadov2010a}
\bibinfo{author}{Chadov, S.} \emph{et~al.}
\newblock \bibinfo{title}{{Tunable multifunctional topological insulators in
  ternary Heusler compounds.}}
\newblock \emph{\bibinfo{journal}{Nat. Mater.}} \textbf{\bibinfo{volume}{9}},
  \bibinfo{pages}{541} (\bibinfo{year}{2010}).
\bibinfo{doi}{DOI: 10.1038/nmat2770}.

\bibitem{Nowak2015}
\bibinfo{author}{Nowak, B.}, \bibinfo{author}{Pavlosiuk, O.} \& \bibinfo{author}{Kaczorowski, D.}
\newblock \bibinfo{title}{{Band Inversion in Topologically Nontrivial Half-Heusler Bismuthides: $^{209}$Bi NMR Study. }}
\newblock \emph{\bibinfo{journal}{J. Phys. Chem. C}}  \textbf{\bibinfo{volume}{119}}, \bibinfo{pages}{2770} (\bibinfo{year}{2015}).
\bibinfo{doi}{DOI: 10.1021/jp5115493}.

\bibitem{Ando2013}
\bibinfo{author}{Ando, Y.}
\newblock \bibinfo{title}{{Topological Insulator Materials}}.
\newblock \emph{\bibinfo{journal}{J. Phys. Soc. Jpn.}}
  \textbf{\bibinfo{volume}{82}}, \bibinfo{pages}{102001}
  (\bibinfo{year}{2013}).
\bibinfo{doi}{DOI: 10.7566/JPSJ.82.102001}.

\bibitem{Hasan2010a}
\bibinfo{author}{Hasan, M.~Z.} \& \bibinfo{author}{Kane, C.~L.}
\newblock \bibinfo{title}{{Colloquium: Topological insulators}}.
\newblock \emph{\bibinfo{journal}{Rev. Mod. Phys.}}
  \textbf{\bibinfo{volume}{82}}, \bibinfo{pages}{3045}
  (\bibinfo{year}{2010}).
\bibinfo{doi}{DOI: 10.1103/RevModPhys.82.3045}.

\bibitem{Vafek2014}
\bibinfo{author}{Vafek, O.} \& \bibinfo{author}{Vishwanath, A.}
\newblock \bibinfo{title}{{ Dirac fermions in solids-from high $T_c$ cuprates
  and graphene to topological insulators and Weyl semimetals.}}
\newblock \emph{\bibinfo{journal}{Ann. Rev. Cond. Matter. Phys.}}
   \textbf{\bibinfo{volume}{5}}, \bibinfo{pages}{83} (\bibinfo{year}{2014}).
\bibinfo{doi}{DOI:~10.1146/annurev-conmatphys-031113-133841}.

\bibitem{Young2012}
\bibinfo{author}{Young, S.~M.} \emph{et~al.}
\newblock \bibinfo{title}{{Dirac Semimetal in Three Dimensions}}.
\newblock \emph{\bibinfo{journal}{Phys. Rev. Lett.}}
  \textbf{\bibinfo{volume}{108}}, \bibinfo{pages}{140405}
  (\bibinfo{year}{2012}).
\bibinfo{doi}{DOI: 10.1103/PhysRevLett.108.140405}.

\bibitem{Castro2009}
\bibinfo{author}{Castro, A.}, \bibinfo{author}{Guinea, F.},
  \bibinfo{author}{Peres, N. M.~R.}, \bibinfo{author}{Novoselov, K.~S.} \&
  \bibinfo{author}{Geim, A.~K.}
\newblock \bibinfo{title}{{The electronic properties of graphene.}}
\newblock \emph{\bibinfo{journal}{Rev. Mod. Phys.}}
  \textbf{\bibinfo{volume}{81}}, \bibinfo{pages}{109}
  (\bibinfo{year}{2002}).
\bibinfo{doi}{DOI: 10.1103/RevModPhys.81.109}.

\bibitem{Wan2011}
\bibinfo{author}{Wan, X.}, \bibinfo{author}{Turner, A.~M.},
  \bibinfo{author}{Vishwanath, A.} \& \bibinfo{author}{Savrasov, S.~Y.}
\newblock \bibinfo{title}{{Topological semimetal and Fermi-arc surface states
  in the electronic structure of pyrochlore iridates}}.
\newblock \emph{\bibinfo{journal}{Phys. Rev. B}} \textbf{\bibinfo{volume}{83}},
  \bibinfo{pages}{205101} (\bibinfo{year}{2011}).
\bibinfo{doi}{DOI: 10.1103/PhysRevB.83.205101}.

\bibitem{Yi2014}
\bibinfo{author}{Yi, H.} \emph{et~al.}
\newblock \bibinfo{title}{{Evidence of Topological Surface State in
  Three-Dimensional Dirac Semimetal Cd$_3$As$_2$.}}
\newblock \emph{\bibinfo{journal}{Sci. Rep.}} \textbf{\bibinfo{volume}{4}},
  \bibinfo{pages}{6106} (\bibinfo{year}{2014}).
\bibinfo{doi}{DOI: 10.1038/srep06106}.

\bibitem{Riedemann1996}
\bibinfo{author}{Riedemann, T.~M.}
\newblock \bibinfo{title}{{Heat capacities, magnetic properties, and
  resistivities of ternary RPdBi alloys where R = La, Nd, Gd, Dy, Er, and Lu}}.
\newblock \bibinfo{type}{MSc thesis}, \bibinfo{school}{Iowa State University},
  \bibinfo{address}{Ames} (\bibinfo{year}{1996}).
\bibinfo{doi}{DOI: 10.2172/251374}.

\bibitem{Gofryk2005}
\bibinfo{author}{Gofryk, K.}, \bibinfo{author}{Kaczorowski, D.},
  \bibinfo{author}{Plackowski, T.}, \bibinfo{author}{Leithe-Jasper, A.} \&
  \bibinfo{author}{Grin, Y.}
\newblock \bibinfo{title}{{Magnetic and transport properties of the
  rare-earth-based Heusler phases RPdZ and RPd$_2$Z (Z=Sb,Bi)}}.
\newblock \emph{\bibinfo{journal}{Phys. Rev. B}} \textbf{\bibinfo{volume}{72}},
  \bibinfo{pages}{094409} (\bibinfo{year}{2005}).
\bibinfo{doi}{DOI: 10.1103/PhysRevB.72.094409}.

\bibitem{Gofryk2011}
\bibinfo{author}{Gofryk, K.}, \bibinfo{author}{Kaczorowski, D.},
  \bibinfo{author}{Plackowski, T.}, \bibinfo{author}{Leithe-Jasper, A.} \&
  \bibinfo{author}{Grin, Y.}
\newblock \bibinfo{title}{{Magnetic and transport properties of
  rare-earth-based half-Heusler phases RPdBi: Prospective systems for
  topological quantum phenomena}}.
\newblock \emph{\bibinfo{journal}{Phys. Rev. B}} \textbf{\bibinfo{volume}{84}},
  \bibinfo{pages}{035208} (\bibinfo{year}{2011}).
\bibinfo{doi}{DOI: 10.1103/PhysRevB.84.035208}.

\bibitem{Kaczorowski2005}
\bibinfo{author}{Kaczorowski, D.}, \bibinfo{author}{Gofryk, K.},
  \bibinfo{author}{Plackowski, T.}, \bibinfo{author}{Leithe-Jasper, A.} \&
  \bibinfo{author}{Grin, Y.}
\newblock \bibinfo{title}{{Unusual features of erbium-based Heusler phases}}.
\newblock \emph{\bibinfo{journal}{J. Magn. Magn. Mater.}}
  \textbf{\bibinfo{volume}{290-291}}, \bibinfo{pages}{573}
  (\bibinfo{year}{2005}).
\bibinfo{doi}{DOI:~10.1016/j.jmmm.2004.11.538}.

\bibitem{Al-Sawai2010}
\bibinfo{author}{Al-Sawai, W.} \emph{et~al.}
\newblock \bibinfo{title}{{Topological electronic structure in half-Heusler
  topological insulators}}.
\newblock \emph{\bibinfo{journal}{Phys. Rev. B}} \textbf{\bibinfo{volume}{82}},
  \bibinfo{pages}{125208} (\bibinfo{year}{2010}).
\bibinfo{doi}{DOI: 10.1103/PhysRevB.82.125208}.

\bibitem{Lin2010}
\bibinfo{author}{Lin, H.} \emph{et~al.}
\newblock \bibinfo{title}{{Half-Heusler ternary compounds as new
  multifunctional experimental platforms for topological quantum phenomena.}}
\newblock \emph{\bibinfo{journal}{Nat. Mater.}} \textbf{\bibinfo{volume}{9}},
  \bibinfo{pages}{546} (\bibinfo{year}{2010}).
\bibinfo{doi}{DOI: 10.1038/nmat2771}.

\bibitem{Goll2008a}
\bibinfo{author}{Goll, G.} \emph{et~al.}
\newblock \bibinfo{title}{{Thermodynamic and transport properties of the
  non-centrosymmetric superconductor LaBiPt}}.
\newblock \emph{\bibinfo{journal}{Physica B}}
  \textbf{\bibinfo{volume}{403}}, \bibinfo{pages}{1065}
  (\bibinfo{year}{2008}).
\bibinfo{doi}{DOI: 10.1016/j.physb.2007.10.089}.

\bibitem{Tafti2013}
\bibinfo{author}{Tafti, F.~F.} \emph{et~al.}
\newblock \bibinfo{title}{{Superconductivity in the noncentrosymmetric
  half-Heusler compound LuPtBi: A candidate for topological
  superconductivity}}.
\newblock \emph{\bibinfo{journal}{Phys. Rev. B}} \textbf{\bibinfo{volume}{87}},
  \bibinfo{pages}{184504} (\bibinfo{year}{2013}).
\bibinfo{doi}{DOI: 10.1103/PhysRevB.87.184504}.

\bibitem{Butch2011a}
\bibinfo{author}{Butch, N.~P.}, \bibinfo{author}{Syers, P.},
  \bibinfo{author}{Kirshenbaum, K.}, \bibinfo{author}{Hope, A.~P.} \&
  \bibinfo{author}{Paglione, J.}
\newblock \bibinfo{title}{{Superconductivity in the topological semimetal
  YPtBi}}.
\newblock \emph{\bibinfo{journal}{Phys. Rev. B}} \textbf{\bibinfo{volume}{84}},
  \bibinfo{pages}{220504} (\bibinfo{year}{2011}).
\bibinfo{doi}{DOI: 10.1103/PhysRevB.84.220504}.

\bibitem{Bay2012}
\bibinfo{author}{Bay, T.}, \bibinfo{author}{Naka, T.},
  \bibinfo{author}{Huang, Y.} \& \bibinfo{author}{de Visser, A.}
\newblock \bibinfo{title}{{Superconductivity in noncentrosymmetric YPtBi under pressure}}.
\newblock \emph{\bibinfo{journal}{Phys. Rev. B}} \textbf{\bibinfo{volume}{86}},
  \bibinfo{pages}{064515} (\bibinfo{year}{2012}).
\bibinfo{doi}{DOI: 10.1103/PhysRevB.86.064515}.

\bibitem{Xu2014a}
\bibinfo{author}{Xu, G.} \emph{et~al.}
\newblock \bibinfo{title}{{Weak Antilocalization Effect and Noncentrosymmetric
  Superconductivity in a Topologically Nontrivial Semimetal LuPdBi.}}
\newblock \emph{\bibinfo{journal}{Sci. Rep.}} \textbf{\bibinfo{volume}{4}},
  \bibinfo{pages}{5709} (\bibinfo{year}{2014}).
\bibinfo{doi}{DOI: 10.1038/srep05709}.

\bibitem{Pavlosiuk2015}
\bibinfo{author}{Pavlosiuk, O.}, \bibinfo{author}{Kaczorowski, D.} \&
  \bibinfo{author}{Wi\'{s}niewski, P.}
\newblock \bibinfo{title}{{Shubnikov-de Haas oscillations, weak
  antilocalization effect and large linear magnetoresistance in the putative
  topological superconductor LuPdBi.}}
\newblock \emph{\bibinfo{journal}{Sci. Rep.}} \textbf{\bibinfo{volume}{5}},
  \bibinfo{pages}{9158} (\bibinfo{year}{2015}).
\bibinfo{doi}{DOI: 10.1038/srep09158}.

\bibitem{Fu2008}
\bibinfo{author}{Fu, L.} \& \bibinfo{author}{Kane, C.}
\newblock \bibinfo{title}{{Superconducting Proximity Effect and Majorana
  Fermions at the Surface of a Topological Insulator}}.\newblock \emph{\bibinfo{journal}{Phys. Rev. Lett.}} \textbf{\bibinfo{volume}{100}}, \bibinfo{pages}{096407} (\bibinfo{year}{2008}).
\bibinfo{doi}{DOI: 10.1103/PhysRevLett.100.096407}.

\bibitem{Mong2010}
\bibinfo{author}{Mong, R.~S.~K}, \bibinfo{author}{Essin, A.~M} \& \bibinfo{author}{Moore, J.~E}
 \newblock \bibinfo{title}{{Antiferromagnetic topological insulators}}.
\newblock \emph{\bibinfo{journal}{Phys. Rev. B}} \textbf{\bibinfo{volume}{81}},
  \bibinfo{pages}{245209} (\bibinfo{year}{2010}).
 \bibinfo{doi}{DOI: 10.1103/PhysRevB.81.245209}.

\bibitem{Muller2014}
\bibinfo{author}{M$\ddot{\rm u}$ller, R.~A} \emph{et~al.}
 \newblock \bibinfo{title}{{Magnetic structure of GdBiPt: A candidate antiferromagnetic topological insulator}}.
\newblock \emph{\bibinfo{journal}{Phys. Rev. B}} \textbf{\bibinfo{volume}{90}},
  \bibinfo{pages}{041109(R)} (\bibinfo{year}{2014}).
 \bibinfo{doi}{DOI: 10.1103/PhysRevB.90.041109}.

\bibitem{Pavlosiuk2015a}
\bibinfo{author}{Pavlosiuk, O.}, \bibinfo{author}{Filar, K.},
  \bibinfo{author}{Wi\'{s}niewski, P.} \& \bibinfo{author}{Kaczorowski, D.}
\newblock \bibinfo{title}{{Magnetic order and SdH effect in half-Heusler phase
  ErPdBi}}.
\newblock \emph{\bibinfo{journal}{Acta Phys. Pol. A}}
  \textbf{\bibinfo{volume}{127}}, \bibinfo{pages}{656}
  (\bibinfo{year}{2015}).
\bibinfo{doi}{DOI:~10.12693/APhysPolA.127.656}.

\bibitem{rhoCEF1963}
\bibinfo{author}{Van Peski-Tinbergen, T.} \& \bibinfo{author}{A.J. Dekker}
\newblock \bibinfo{title}{{Spin-dependent scattering and resistivity of magnetic metals and alloys}}.
\newblock \emph{\bibinfo{journal}{Physica}} \textbf{\bibinfo{volume}{29}},  \bibinfo{pages}{917} (\bibinfo{year}{1963}).
\bibinfo{doi}{DOI: 10.1016/S0031-8914(63)80182-2}.

\bibitem{rhoCEF1980}
\bibinfo{author}{Hessel Andersen N.}, \bibinfo{author}{Jensen, J.}, \bibinfo{author}{Smith, H.}, \bibinfo{author}{Splittorff, O.} \& \bibinfo{author}{Vogt, O.} \newblock \bibinfo{title}{{Electrical resistivity of the singlet-ground-state system Tb$_c$Y$_{1-c}$Sb}}. \newblock \emph{\bibinfo{journal}{Phys. Rev. B}} \textbf{\bibinfo{volume}{21}}, \bibinfo{pages}{189} (\bibinfo{year}{1980}).
\bibinfo{doi}{DOI: 10.1103/PhysRevB.21.189}.

\bibitem{Krusius1969}
\bibinfo{author}{Krusius M.} \bibinfo{author}{Anderson A. C.} \& \bibinfo{author}{Holmstr\"{o}m, B.}
\newblock \bibinfo{title}{{Calorimetric investigation of hyperfine interactions in metallic Ho and Tb}}.
\newblock \emph{\bibinfo{journal}{Phys. Rev.}} \textbf{\bibinfo{volume}{177}},
 \bibinfo{pages}{910} (\bibinfo{year}{1969}).
 \bibinfo{doi}{DOI: 10.1103/PhysRev.177.910}.

\bibitem{Rapp1999}
\bibinfo{author}{Rapp R.} \& \bibinfo{author}{ Massalami M.}
\newblock \bibinfo{title}{{Nonsuperconductivity and magnetic features of the intermetallic borocarbide HoCo$_2$B$_2$C}}.
\newblock \emph{\bibinfo{journal}{Phys. Rev. B}} \textbf{\bibinfo{volume}{60}},
  \bibinfo{pages}{3355} (\bibinfo{year}{1999}).
\bibinfo{doi}{DOI: 10.1103/PhysRevB.60.3355}.

\bibitem{Tari2003}
\bibinfo{author}{Tari, A.}
\newblock \emph{\bibinfo{title}{{The Specific Heat of Matter at Low Temperatures}}}
  (\bibinfo{publisher}{Imperial College Press},
  \bibinfo{address}{London}, \bibinfo{year}{2003}).
\bibinfo{doi}{DOI: 10.1142/9781860949395}.

\bibitem{Lea1962}
\bibinfo{author}{Lea, K. R} \bibinfo{author}{Leask, M. J. M.} \&
  \bibinfo{author}{Wolf, P. F.}
\newblock \bibinfo{title}{{The raising of angular momentum degeneracy of $f$-electron terms by cubic crystal fields}}.
\newblock \emph{\bibinfo{journal}{J. Phys. Chem. Solids}} \textbf{\bibinfo{volume}{23}}, \bibinfo{pages}{1381} (\bibinfo{year}{1962}).
\bibinfo{doi}{DOI:~10.1016/0022-3697(62)90192-0}.

\bibitem{Karla1999}
\bibinfo{author}{Karla, I.}, \bibinfo{author}{Pierre, J.}, \bibinfo{author}{Murani, A.} \& \bibinfo{author}{Neumann, M.} \newblock \bibinfo{title}{{Crystalline electric field in RNiSb compounds investigated by inelastic neutron scattering.}} 
\newblock \emph{\bibinfo{journal}{Physica B}} \textbf{\bibinfo{volume}{271}}, \bibinfo{pages}{294} (\bibinfo{year}{1999}).
\bibinfo{doi}{DOI: 10.1016/S0921-4526(99)00196-9}.

\bibitem{Taskin2012}
\bibinfo{author}{Taskin, A.~A}, \bibinfo{author}{Sasaki, S.}, \bibinfo{author}{Segawa, K.} \& \bibinfo{author}{ Ando, Y.}
\newblock \bibinfo{title}{{Manifestation of Topological Protection in Transport Properties of Epitaxial Bi$_2$Se$_3$ Thin Films}}.
\newblock \emph{\bibinfo{journal}{Phys. Rev. Lett.}}
  \textbf{\bibinfo{volume}{109}}, \bibinfo{pages}{066803}
  (\bibinfo{year}{2012}).
\bibinfo{doi}{DOI: 10.1103/PhysRevLett.109.066803}.

\bibitem{Yang2015}
\bibinfo{author}{Yang, X.}, \bibinfo{author}{Li, Y.}, \bibinfo{author}{Wang, Z.}, \bibinfo{author}{Zhen, Y.} \& \bibinfo{author}{Xu, Z.},
\newblock \bibinfo{title}{{Observation of Negative Magnetoresistance and nontrivial $\pi$ Berry's phase in 3D Weyl semi-metal NbAs}}.
(\bibinfo{year}{2015}).
\newblock \eprint [arXiv]{preprint arXiv:1506.02283v1}.

\bibitem{Hikami1980}
\bibinfo{author}{Hikami, S.}, \bibinfo{author}{Larkin, A.~I.} \&
  \bibinfo{author}{Nagaoka, Y.}
\newblock \bibinfo{title}{{Spin-Orbit Interaction and Magnetoresistance in the
  Two Dimensional Random System}}.
\newblock \emph{\bibinfo{journal}{Prog. Theor. Phys.}}
  \textbf{\bibinfo{volume}{63}}, \bibinfo{pages}{707}
  (\bibinfo{year}{1980}).
\bibinfo{doi}{DOI:~10.1143/PTP.63.707}.

\bibitem{Sacksteder2014}
\bibinfo{author}{Sacksteder, V.~E.}, \bibinfo{author}{Arnardottir, K.~B.}, \bibinfo{author}{Kettemann, S.} \&  \bibinfo{author}{Shelykh, I.~A.}
\newblock \bibinfo{title}{{Topological effects on the magnetoconductivity in topological insulators}}.
\newblock \emph{\bibinfo{journal}{Phys. Rev. B}}
  \textbf{\bibinfo{volume}{90}}, \bibinfo{pages}{235148}
  (\bibinfo{year}{2014}).
\bibinfo{doi}{DOI:~10.1103/PhysRevB.90.235148}.

\bibitem{Gofryk2007}
\bibinfo{author}{Gofryk, K.} \emph{et~al.}
\newblock \bibinfo{title}{{Magnetic, transport, and thermal properties of the half-Heusler compounds ErPdSb and YPdSb}}.
\newblock \emph{\bibinfo{journal}{Phys. Rev. B}} \textbf{\bibinfo{volume}{75}},
  \bibinfo{pages}{224426} (\bibinfo{year}{2007}).
\bibinfo{doi}{DOI: 10.1103/PhysRevB.75.224426}.

\bibitem{Mun2013}
\bibinfo{author}{Mun, E.} \emph{et~al.}
\newblock \bibinfo{title}{{Magnetic-field-tuned quantum criticality of the heavy-fermion system YbPtBi}}.
\newblock \emph{\bibinfo{journal}{Phys. Rev. B}} \textbf{\bibinfo{volume}{87}},
  \bibinfo{pages}{075120} (\bibinfo{year}{2013}).
\bibinfo{doi}{DOI: 10.1103/PhysRevB.87.075120}.

\bibitem{Karla1998}
\bibinfo{author}{Karla, I.}, \bibinfo{author}{Pierre, J.}, \& \bibinfo{author}{Skolozdra, R.}
\newblock \bibinfo{title}{{Physical properties and giant magnetoresistance in RNiSb compounds. }}
\newblock \emph{\bibinfo{journal}{J. Alloys Compds.}}
\textbf{\bibinfo{volume}{265}}, \bibinfo{pages}{42}  (\bibinfo{year}{1998}).
\bibinfo{doi}{DOI: 10.1016/S0925-8388(97)00419-2}

\bibitem{Takita1974}
\bibinfo{author}{Takita, K.}, \bibinfo{author}{Tanimura, N.} \&
  \bibinfo{author}{Tanaka, S.}
\newblock \bibinfo{title}{{Anomalous Magnetoresistance Effect in HgTe under
  Uniaxial Compression at Low Temperatures}}.
\newblock In \emph{\bibinfo{booktitle}{Proc. 12th Int. Conf. Phys. Semicond.}},
  \bibinfo{pages}{1152} (\bibinfo{publisher}{Vieweg+Teubner Verlag},
  \bibinfo{address}{Wiesbaden}, \bibinfo{year}{1974}).
\bibinfo{doi}{DOI: 10.1007/978-3-322-94774-1\_199}.

\bibitem{Hamlin2012}
\bibinfo{author}{Hamlin, J.~J.} \emph{et~al.}
\newblock \bibinfo{title}{{High pressure transport properties of the topological insulator Bi$_2$Se$_3$}}.
\newblock \emph{\bibinfo{journal}{J. Phys.: Condens. Matter}} \textbf{\bibinfo{volume}{24}}, \bibinfo{pages}{035602} (\bibinfo{year}{2012}).
\bibinfo{doi}{DOI: 10.1088/0953-8984/24/3/035602}.

\bibitem{Klier2015}
\bibinfo{author}{Klier, J.}, \bibinfo{author}{Gornyi, I.~V.} \& \bibinfo{author}{Mirlin, A.~D.}
\newblock \bibinfo{title}{{Transversal magnetoresistance in Weyl semimetals}}.
(\bibinfo{year}{2015}).
\newblock \eprint [arXiv]{preprint arXiv:1507.03481v1}.

\bibitem{Potter2014a}
\bibinfo{author}{Potter, A.~C.}, \bibinfo{author}{Kimchi, I.} \&
  \bibinfo{author}{Vishwanath, A.}
\newblock \bibinfo{title}{{Quantum oscillations from surface Fermi arcs in Weyl
  and Dirac semimetals}}.
\newblock \emph{\bibinfo{journal}{Nat. Commun.}} \textbf{\bibinfo{volume}{5}},
  \bibinfo{pages}{1} (\bibinfo{year}{2014}).
\bibinfo{doi}{DOI: 10.1038/ncomms6161}.

\bibitem{Shoenberg1984}
\bibinfo{author}{Shoenberg, D.}
\newblock \emph{\bibinfo{title}{{Magnetic Oscillations in Metals}}}
  (\bibinfo{publisher}{Cambridge University Press},
  \bibinfo{address}{Cambridge}, \bibinfo{year}{1984}).
\bibinfo{doi}{DOI: 10.1017/CBO9780511897870}.

\bibitem{Lukyanchuk2004}
\bibinfo{author}{Luk'yanchuk, I.~A.} \&
  \bibinfo{author}{Kopelevich, Y.}
\newblock \bibinfo{title}{{Phase analysis of quantum oscillations in graphite}}.
\newblock \emph{\bibinfo{journal}{Phys. Rev. Lett.}} \textbf{\bibinfo{volume}{93}},
  \bibinfo{pages}{166402} (\bibinfo{year}{2004}).
\bibinfo{doi}{DOI: 10.1103/PhysRevLett.93.166402}.

\bibitem{Wang2013}
\bibinfo{author}{Wang, W.} \emph{et~al.}
\newblock \bibinfo{title}{{Large Linear Magnetoresistance and Shubnikov-de Hass
  Oscillations in Single Crystals of YPdBi Heusler Topological Insulators.}}
\newblock \emph{\bibinfo{journal}{Sci. Rep.}} \textbf{\bibinfo{volume}{3}},
  \bibinfo{pages}{2181} (\bibinfo{year}{2013}).
\bibinfo{doi}{DOI: 10.1038/srep02181}.

\bibitem{Wosnitza2006a}
\bibinfo{author}{Wosnitza, J.} \emph{et~al.}
\newblock \bibinfo{title}{{Magnetic-field- and temperature-dependent Fermi surface of CeBiPt}}.
\newblock \emph{\bibinfo{journal}{New J. Phys.}} \textbf{\bibinfo{volume}{8}},
  \bibinfo{pages}{174} (\bibinfo{year}{2006}).
\bibinfo{doi}{DOI: 10.1088/1367-2630/8/9/174}.

\bibitem{Marazza1980}
\bibinfo{author}{Marazza, R., Rossi, D. \& Ferro, R.}
\newblock \bibinfo{title}{{MgAgAs-type phases in the ternary systems of rare-earths with palladium and bismuth}}. 
\newblock \emph{\bibinfo{journal}{Gazz. Chim. Ital.}}  \textbf{\bibinfo{volume}{110}},
 \bibinfo{pages}{357} (\bibinfo{year}{1980}). 

\bibitem{Gukasov2013}
\bibinfo{author}{Gukasov, A.} \emph{et~al.}
\newblock \bibinfo{title}{{Very Intense Polarized (VIP) Neutron Diffractometer at the ORPHEE Reactor in Saclay.}}
\newblock \emph{\bibinfo{journal}{Phys. Procedia}} \textbf{\bibinfo{volume}{42}},  \bibinfo{pages}{150} (\bibinfo{year}{2013}).
\bibinfo{doi}{DOI: 10.1016/j.phpro.2013.03.189}.

\end{thebibliography}

\begin{thebibliography}{100}
\expandafter\ifx\csname url\endcsname\relax
  \def\url#1{\texttt{#1}}\fi
\expandafter\ifx\csname urlprefix\endcsname\relax\def\urlprefix{URL }\fi
\providecommand{\bibinfo}[2]{#2}
\providecommand{\eprint}[2][]{\url{#2}}

\bibitem{Marazza1980}
\bibinfo{author}{Marazza, R., Rossi, D. \& Ferro, R.}
\newblock \bibinfo{title}{{MgAgAs-type phases in the ternary systems of rare-earths with palladium and bismuth}}. 
\newblock \emph{\bibinfo{journal}{Gazz. Chim. Ital.}}  \textbf{\bibinfo{volume}{110}},  \bibinfo{pages}{357} (\bibinfo{year}{1980}). 

\bibitem{Nikitin2015}
\bibinfo{author}{Nikitin, A.~M.} \emph{et~al.}
\newblock \bibinfo{title}{{Magnetic and superconducting phase diagram of the
  half-Heusler topological semimetal HoPdBi}}.
\newblock \emph{\bibinfo{journal}{J. Phys. Condens. Matter}}
  \textbf{\bibinfo{volume}{27}}, \bibinfo{pages}{275701}
  (\bibinfo{year}{2015}).

\bibitem{Karla1998}
\bibinfo{author}{Karla, I.}, \bibinfo{author}{Pierre, J.}, \& \bibinfo{author}{Skolozdra, R.}
\newblock \bibinfo{title}{{Physical properties and giant magnetoresistance in RNiSb compounds. }}
\newblock \emph{\bibinfo{journal}{J. Alloys Compds.}}
\textbf{\bibinfo{volume}{265}}, \bibinfo{pages}{42}  (\bibinfo{year}{1998}).

\bibitem{Nakajima2015}
\bibinfo{author}{Nakajima, Y.} \emph{et~al.}
\newblock \bibinfo{title}{{Topological RPdBi half-Heusler semimetals: A new
  family of noncentrosymmetric magnetic superconductors}}.
\newblock \emph{\bibinfo{journal}{Sci. Adv.}} \textbf{\bibinfo{volume}{1}},
  \bibinfo{pages}{e1500242} (\bibinfo{year}{2015}).
\end{thebibliography}
\end{document}